\newcommand{\be}{\begin{equation}}
\newcommand{\ee}{\end{equation}}

\documentclass[journal,10pt]{IEEEtran}

\usepackage{booktabs}
\usepackage{psfrag}
\ifCLASSINFOpdf
   \usepackage[pdftex]{graphicx}

\else
  
   \usepackage[dvips]{graphicx}

\fi
\usepackage[cmex10]{amsmath}
\usepackage{amsthm,epstopdf,nicefrac,pgfplots}

  \pgfplotsset{compat=newest} 
  \pgfplotsset{plot coordinates/math parser=false}

\usepackage[bookmarks=false]{hyperref}

\usepackage{rotating}

\usepackage{mathtools}

\usepackage{gensymb}

\newcommand\copyrighttext{%
  \footnotesize \textcopyright 2015 IEEE.  Personal use of this material is permitted. Permission from IEEE must be obtained for all other uses, in any current or future media, including reprinting/republishing this material for advertising or promotional purposes, creating new collective works, for resale or redistribution to servers or lists, or reuse of any copyrighted component of this work in other works. 
 DOI: \href{https://doi.org/10.1109/TIM.2015.2463331}{10.1109/TIM.2015.2463331}
 }
\newcommand\copyrightnotice{%
\begin{tikzpicture}[remember picture,overlay]
\node[anchor=south,yshift=10pt] at (current page.south) {\fbox{\parbox{\dimexpr\textwidth-\fboxsep-\fboxrule\relax}{\copyrighttext}}};
\end{tikzpicture}%
}

\renewcommand\appendix{\par
  \setcounter{section}{0}
  \setcounter{subsection}{0}
  \setcounter{figure}{0}
  \setcounter{table}{0}
  \renewcommand\thesection{Appendix \Alph{section}}
  \renewcommand\thefigure{\Alph{section}\arabic{figure}}
  \renewcommand\thetable{\Alph{section}\arabic{table}}
}

\usepackage{color}

\twocolumn

  \newlength\fheight 
    \newlength\fwidth 
\usepackage{balance}
\usepackage{algorithm}
\usepackage{algpseudocode}

\begin{document}
\title{Accurate Sine Wave Amplitude Measurements\\ using Nonlinearly Quantized Data}

%\newtheorem{theorem}{Theorem}[section]
%\newtheorem{lemma}[theorem]{Lemma}

\author{P.~Carbone,~\IEEEmembership{Fellow Member,~IEEE}\thanks{P. Carbone and Antonio Moschitta are with the University of Perugia - Engineering Department, via G. Duranti, 93 - 06125 Perugia Italy,}
and~J.~Schoukens,~\IEEEmembership{Fellow Member,~IEEE}\thanks{J. Schoukens is with the Vrije Universiteit Brussel, Department ELEC, Pleinlaan 2, B1050 Brussels, Belgium.}
and~I.~Kollar,~\IEEEmembership{Fellow Member,~IEEE}\thanks{I. Kollar is with the Budapest University of Technology and Economics, Department of Measurement and Information Systems, 1521 Budapest, Hungary.}
and~A.~Moschitta~\IEEEmembership{Member,~IEEE}}

\maketitle
\copyrightnotice
\begin{abstract}
\boldmath
%\today \newline
The estimation of the amplitude of a 
sine wave from the sequence 
of its quantized samples is a typical problem 
in instrumentation and measurement.
A standard approach for its 
solution makes use 
of a least squares estimator (LSE)
that, however, does not perform optimally  
in the presence of quantization errors. 
In fact, if the quantization error 
cannot be modeled as an additive {\em noise source},  
as it often happens in practice,
the LSE returns biased estimates. 

In this paper we consider the estimation of the amplitude of a noisy sine wave
after quantization. The proposed technique
is based on a uniform distribution of signal phases
% does not require synchronous or coherent
%sampling, nor that 
and it does not require that the quantizer has equally spaced transition levels. Experimental results show that this technique removes the estimation 
bias associated to the usage of the LSE and that it is sufficiently 
robust with respect to small uncertainties in the known values of transition levels.

\end{abstract}

\begin{IEEEkeywords}
Quantization, estimation, nonlinear estimation problems, identification, nonlinear quantizers.
\end{IEEEkeywords}

\newcommand{\fg}[1]{{\frac{1}{\sqrt{2\pi}\sigma} e^{-\frac{{#1}^2}{2\sigma^2}}}} 

% \fi
\IEEEpeerreviewmaketitle

%\balance

\section{Introduction}
Almost all modern instruments 
acquire data by means of Analog--to--Digital Converters (ADCs).  
While technology has progressed to yield ADCs with increasing performance 
in terms of power consumption, effective bits and 
rate--of--convergence, 
the nonlinear transformation implied by the conversion process may still 
result in inaccurate {\color{black} estimates} of input signal parameters.
In fact, the majority of results associated to the {\em quantization} 
operation performed by ADCs, are derived by assuming 
a perfectly uniform input--output characteristic, 
with equally spaced transition levels.  
When this occurs, the ADC is termed {\em linear} with an obvious
semantic  abuse, given that even a uniformly spaced stepwise input--output characteristic results in a nonlinear transformation of the input signal.
Based on these hypotheses, general properties of quantized signals are derived that 
refer to the analysis of spectra \cite{Gray}--\nocite{Bennett,Schuchman}\cite{Blachman}, determination of the effect of
{\em dithering} both in amplitude-- and in frequency--domains \cite{Kollarbias}--\nocite{ GrayStockham,Zimmerman}\cite{CarbonePetri}, application of the {quantization theorem} \cite{KollarBook}, analysis of the quantization error probability density function 
%\cite{pdf} 
and estimation of the parameters of a sine wave using its quantized samples \cite{DalletBelegaPetri, PintelonSchoukens}. 

More evolved models recognize that ADCs, in practice, are characterized by non--evenly spaced transition levels whose value may also vary 
depending on usage conditions and environmental factors.  
Fewer results are available on the characteristics of {\em nonlinearly} 
quantized signals. 
{\color{black} As an example, } an interesting area of research in this field 
is represented by methods for the  
%A subject that was addressed by some research groups is that of the 
{\em  
compensation} of  the input--output characteristics 
of an ADC to reduce the effects associated to 
the non--uniform spacing of transition levels \cite{Handel}.

The problem of estimating the amplitude of a sine wave using its 
quantized samples is relevant in several engineering applications: for ADC testing purposes \cite{IEEE1241}{\color{black}\cite{BlairLinnenbrink}}, in the estimation of power quality associated to electrical systems \cite{CristaldiFerreroSalicone}, in the characterization of waveform digitizers \cite{IEEE1057}  {\color{black} and in the measurement of impedances \cite{Ramos},} just to name a few. 
Quantization is always affected by some additive noise contributions. 
The noise may be  artificially added, as when dithering is performed \cite{AD9265},
or just be the effect of input--referred noise sources associated 
to the behavior of electronic devices. 
It is known that small amount of additive noise added before quantization may  
linearize the stepwise input--output characteristics, but that large amount of noise is needed for the linearization of quantizers 
with non--uniformly distributed transition levels {\color{black} \cite{Wagdy}}. 

To solve this identification problem, 
the least square{\color{black}s} estimator (LSE) is often used {\color{black} \cite{Alegria}\cite{HandelLSE}}.
However, the nonlinearity 
renders the estimator progressively more biased and far from optimal, 
as the ADC characteristics increasingly departs from uniformity.
Previous work on this subject was {\color{black} published} in \cite{WangYinZhangZhao},
 \cite{104} where a general 
framework 
for system identification based on nonlinearly quantized data is described. 
In \cite{CarboneSchoukensMoschitta} a similar 
approach was used to provide
estimates of amplitude and initial record 
phase of a {\em synchronously} sampled sine wave. 

In this paper we consider the problem of estimating the amplitude of a  
noisy sine wave by using quantized samples. 
Data are considered quantized by a device having known transition levels 
that are not necessarily uniformly spaced in the signal input range. 
It will be shown at first that the LSE fails to provide unbiased 
estimates. 
Then a new estimator is proposed that does not require 
coherent sampling nor knowledge
of initial record phase as required by the estimator presented in
 \cite{CarboneSchoukensMoschitta}. 
Simulation and experimental results will be used to show that the new estimator:
\begin{itemize}
\item removes most of the bias both when the ADC has uniform or non--uniform transition levels; 
\item is capable to estimate the sine wave amplitude even under severe quantization (e.g., with a 2 bit ADC) and thus,
\end{itemize} 
outperforms the LSE.

\section{Signals and systems}

\begin{figure}
\begin{center}
\includegraphics[scale=0.65]{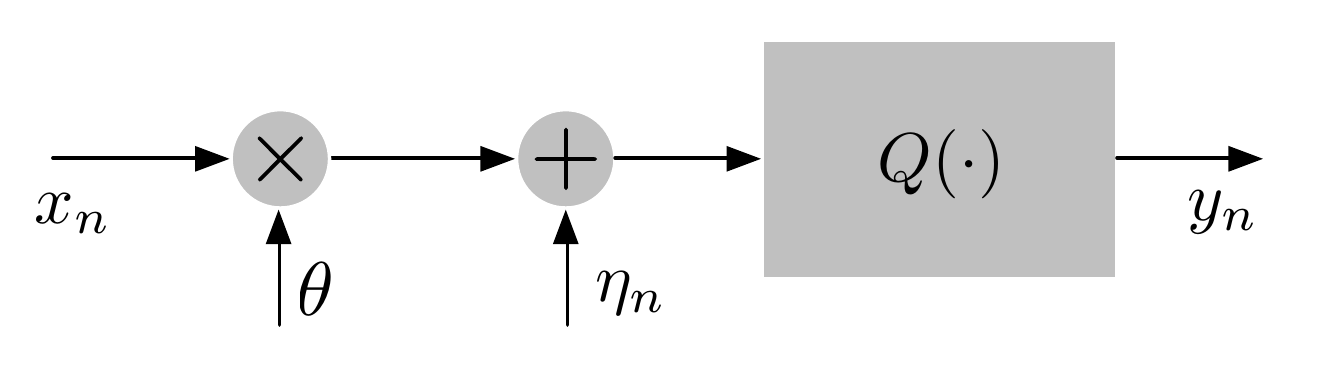}
\caption{The signal chain considered in this paper.\label{signals}}
\end{center}
\end{figure}  

The signal chain considered in this paper is shown in Fig.~\ref{signals}. 
In this figure, 
\be
	x_{n} =  \sin(2\pi \lambda n+{\color{black} \phi_0}) \qquad n=0, \ldots, N-1
\ee
%$\theta_1$ 
represents a known discrete time deterministic sequence 
%known up to a scalar parameter $\theta$, 
with ${\color{black} \phi_0}$ as a possibly unknown initial record phase, not to be estimated,
$n$ the time index and {\color{black} $\lambda= \frac{f}{f_s}$, the normalized sine wave frequency, where $f$ is the sine wave frequency and
$f_s$ is the sampling rate.} 
Moreover, in Fig.~\ref{signals}, $\theta$ represents the constant 
sine wave amplitude to be estimated,
$\eta_n$ is a zero--mean noise sequence with known 
{\color{black} probability density function (PDF)}
and {\color{black} statistically} independent outcomes. Observe that in this case a basic assumption is that the DC level is exactly zero.

In Fig.~\ref{signals}, $Q(\cdot)$ models the instantaneous effect of the ADC on the signal. It  is characterized by $(L-1)$ known transition levels that do not need  
to be uniformly spaced in the input range, given by the normalized interval $[-1,1]$.
%while $y[\cdot]$ represents the sequence of output data. 
If some of the assumed parameters are unknown, e.g. noise variance or threshold levels, 
they need to be estimated during an initial system calibration phase.
Each record{\color{black}, obtained by collecting quantizer output data,} contains $N$ samples $y_n$, $n=0, \ldots, N-1$ that are processed by the algorithms analyzed in this paper. 

Each ADC output sample can accordingly be modeled as a random variable taking values in $L$ possible categories with probability determined by 
the input sequence, the noise PDF and the ADC transition levels.
Assume also that the quantizer output is equal to
$$
	 y[k]  \coloneqq -\left( \frac{L}{2}-1\right)\Delta+
	 k\Delta, \qquad k=0, \ldots, L-1,
$$
when its input takes values in the interval $[T_{k}, T_{k+1})$, 
where $L=2^b$, $b$ is the number of bits, 
$T_k$ is the  $k$--th quantizer transition level
{\color{black}and $\Delta$ is the quantization step}.
{\color{black}Observe that}
 if a uniform ADC is considered,
$T_k=-\left( \frac{L-1}{2}\right) \Delta +k\Delta$.
Accordingly, $k=0$ and $k=L-1$ correspond to the quantizer output being equal to $-\left(\frac{L}{2}-1 \right) \Delta$ and $\frac{L\Delta}{2}$ respectively.
Also define the quantization error $e_n = y_n-\theta x_n$ and consider as negligible the probability that the quantizer input takes values outside the interval $[-1,1]$. When 
{\color{black} the noise standard deviation
$\sigma=0$, this occurs when the sine wave 
amplitude obeys the bound,} 
$\theta<\left( \frac{L-1}{2}\right) \Delta$. 

\subsection{Problem Statement \label{ps}}
With the signals and systems defined above, the estimation problem can be set as follows:
estimate the sine wave amplitude $\theta$ using an $N$--length 
sequence of samples
{\color{black} obtained by quantizing a noisy version of the
sinusoidal signal.}

\subsection{Problem Analysis}

\setlength\fheight{4.5cm} 
\setlength\fwidth{\fheight*\real{1.55}}

\begin{figure}
%\scalebox{1}{
% Minimal pgfplots version: 1.3
%  http://www.mathworks.com/matlabcentral/fileexchange/22022-matlab2tikz
\begin{tikzpicture}

\begin{axis}[%
width=\fwidth,
height=0.618034\fheight,
at={(0\fwidth,0\fheight)},
scale only axis,
separate axis lines,
every outer x axis line/.append style={black},
every x tick label/.append style={font=\color{black}},
xmin=0.1,
xmax=1,
xlabel={$\theta$},
every outer y axis line/.append style={black},
every y tick label/.append style={font=\color{black}},
ymin=-0.05,
ymax=0.6,
ylabel={$\mbox{estimator bias}/\Delta$}
]
\addplot [color=black,solid,line width=1.3pt,forget plot]
  table[row sep=crcr]{%
0.1	-0.00176417652705396\\
0.11	-0.00322414527818182\\
0.12	-0.00377044383233027\\
0.13	-0.00131393390064716\\
0.14	0.00254526456583903\\
0.15	0.00537563193982749\\
0.16	0.00436456665499918\\
0.17	0.00289824059437649\\
0.18	0.0012938382240435\\
0.19	-0.00453809190183563\\
0.2	-0.00203562794251866\\
0.21	-0.00251618175192903\\
0.22	0.00145111160598788\\
0.23	0.00238420632939551\\
0.24	0.00342243476146109\\
0.25	0.00107741020977414\\
0.26	-0.000556631165409271\\
0.27	-0.0020523060729829\\
0.28	-0.00468290997935128\\
0.29	-0.000987165194686668\\
0.3	-0.00109232793070646\\
0.31	0.00302592529206436\\
0.32	0.00285074283780773\\
0.33	0.00250056920231145\\
0.34	-0.00121650064639311\\
0.35	-0.00187860881865731\\
0.36	-0.00158753576312165\\
0.37	-0.00087861370968767\\
0.38	-0.00158866048573714\\
0.39	0.00168182657066041\\
0.4	0.000819984836255117\\
0.41	0.00176147215313449\\
0.42	0.00191100211299045\\
0.43	0.00147108099000093\\
0.44	-0.000224081545383115\\
0.45	-0.00104190496114143\\
0.46	-0.000732750892211698\\
0.47	-0.00144064966235646\\
0.48	0.00171782731982262\\
0.49	0.00263579049558871\\
0.5	0.000180274515969359\\
0.51	0.00093067562198712\\
0.52	-0.00275850814210798\\
0.53	-0.00521019185862315\\
0.54	0.000232960008645478\\
0.55	0.000467712303475309\\
0.56	0.00170772515184581\\
0.57	0.00319797901789798\\
0.58	0.00216239805894247\\
0.59	0.000942185385611083\\
0.6	-0.000791343047239934\\
0.61	-0.000162917204079349\\
0.62	-0.00115283687222245\\
0.63	-0.000496354904612417\\
0.64	-0.00145976790565783\\
0.65	0.0045961858286887\\
0.66	-0.00119135924421698\\
0.67	0.00143728003519072\\
0.68	-0.00306583951601169\\
0.69	-0.00569885184586383\\
0.7	-0.00024903528287723\\
0.71	-0.0028389091607437\\
0.72	0.0020025198639928\\
0.73	0.000223531292022017\\
0.74	0.00349638610248348\\
0.75	0.00312397470094083\\
0.76	0.000482852356412877\\
0.77	0.00231234547004533\\
0.78	0.00231287314505835\\
0.79	-0.000623639445507251\\
0.8	-0.0019775802140316\\
0.81	0.000665247254062251\\
0.82	0.00264662996136167\\
0.83	0.00309800424770401\\
0.84	-0.00208563431920084\\
0.85	0.00183760309249692\\
0.86	-0.00452889333075746\\
0.87	0.00281179739590698\\
0.88	-0.00269486446518385\\
0.89	-0.00164202002105185\\
0.9	0.00415242318916853\\
0.91	0.0016947088936945\\
0.92	-0.00226368916230513\\
0.93	-0.00304449424237418\\
0.94	0.000264203387644102\\
0.95	0.00162899472962863\\
0.96	0.00111135015049513\\
0.97	0.00241897921949885\\
0.98	0.00281545488616075\\
0.99	0.00161313360382564\\
1	-0.0272714854040714\\
};
\node[right, align=left, inner sep=0mm, font=\fontsize{10pt}{1em}\selectfont, text=black]
at (axis cs:0.85,0.5,0) {$(a)$};
\end{axis}
\end{tikzpicture}%
%\hskip-1.6cm
\begin{minipage}{8cm}
\hskip-0.3cm
\input{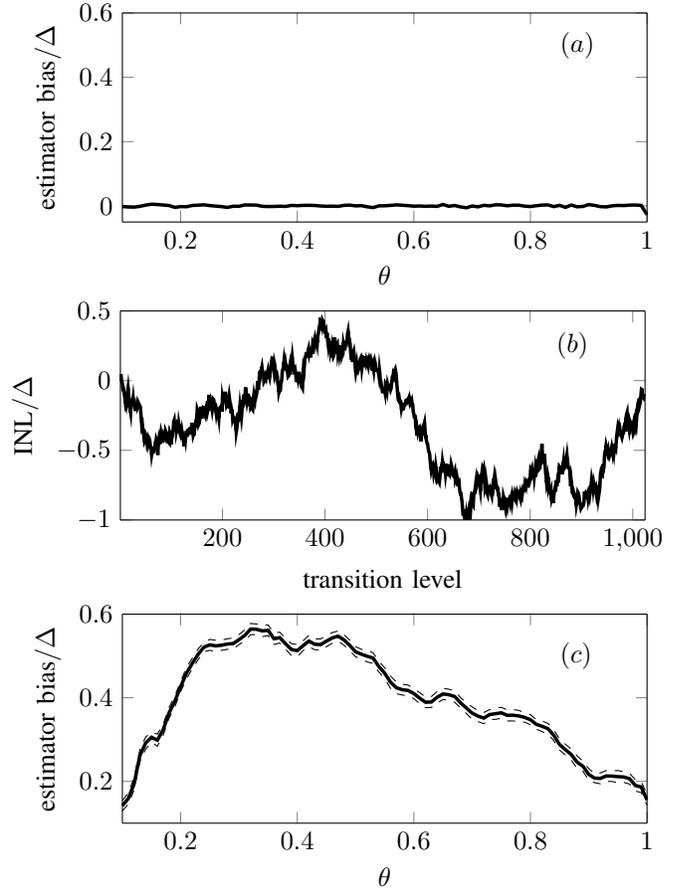}
\end{minipage}
% Minimal pgfplots version: 1.3
%  http://www.mathworks.com/matlabcentral/fileexchange/22022-matlab2tikz
\begin{tikzpicture}

\begin{axis}[%
width=\fwidth,
height=0.618034\fheight,
at={(0\fwidth,0\fheight)},
scale only axis,
separate axis lines,
every outer x axis line/.append style={black},
every x tick label/.append style={font=\color{black}},
xmin=0.1,
xmax=1,
xlabel={$\theta$},
every outer y axis line/.append style={black},
every y tick label/.append style={font=\color{black}},
ymin=0.1,
ymax=0.6,
ylabel={$\mbox{estimator bias}/\Delta$}
]
\addplot [color=black,solid,line width=1.3pt,forget plot]
  table[row sep=crcr]{%
0.1	0.141635867537531\\
0.11	0.158859035447506\\
0.12	0.19331440765523\\
0.13	0.261609345466411\\
0.14	0.291783603720688\\
0.15	0.305488203661355\\
0.16	0.298527051131714\\
0.17	0.323708416939823\\
0.18	0.363611658403016\\
0.19	0.391432926677595\\
0.2	0.426568550713469\\
0.21	0.459006033171732\\
0.22	0.482212824893139\\
0.23	0.504106379405513\\
0.24	0.521096307028259\\
0.25	0.526556016111385\\
0.26	0.524023036544776\\
0.27	0.525646532858786\\
0.28	0.528414907067031\\
0.29	0.529608541660735\\
0.3	0.538981311559326\\
0.31	0.549082541051916\\
0.32	0.564325837446859\\
0.33	0.56374827273487\\
0.34	0.559806320172783\\
0.35	0.56146635130375\\
0.36	0.541144050141554\\
0.37	0.543828271328721\\
0.38	0.528899711934343\\
0.39	0.51531514838203\\
0.4	0.512563033365751\\
0.41	0.524570636203777\\
0.42	0.535588233071621\\
0.43	0.527977626041121\\
0.44	0.527016430387846\\
0.45	0.535458027172098\\
0.46	0.543056722114443\\
0.47	0.547301982276167\\
0.48	0.538626935077787\\
0.49	0.525321684324695\\
0.5	0.510162874542232\\
0.51	0.504064929526123\\
0.52	0.499258467987545\\
0.53	0.49499662791311\\
0.54	0.4746922662261\\
0.55	0.458707137398164\\
0.56	0.437159336696652\\
0.57	0.423807945542706\\
0.58	0.419378808881106\\
0.59	0.417499489702664\\
0.6	0.410083538018966\\
0.61	0.398922728742605\\
0.62	0.389296466390988\\
0.63	0.389800243271907\\
0.64	0.40120411392877\\
0.65	0.409100753553616\\
0.66	0.407203129017262\\
0.67	0.402956976241455\\
0.68	0.388098245942388\\
0.69	0.375439781460784\\
0.7	0.363277717986534\\
0.71	0.356005931676066\\
0.72	0.351327222814007\\
0.73	0.359839901106625\\
0.74	0.361829681347729\\
0.75	0.364006338026059\\
0.76	0.357947985772171\\
0.77	0.358252095981925\\
0.78	0.356511384687508\\
0.79	0.352993916373066\\
0.8	0.347549767807834\\
0.81	0.336428784873988\\
0.82	0.33168822162304\\
0.83	0.326103147997685\\
0.84	0.310377941822082\\
0.85	0.287898645466157\\
0.86	0.275844103512952\\
0.87	0.264178351494706\\
0.88	0.245930582825395\\
0.89	0.236294599221367\\
0.9	0.216735726538161\\
0.91	0.207675857540607\\
0.92	0.207121986337484\\
0.93	0.212916660324368\\
0.94	0.212536971383088\\
0.95	0.211278289223287\\
0.96	0.210512188661824\\
0.97	0.206348117960147\\
0.98	0.190110964764301\\
0.99	0.186579937005718\\
1	0.15546558500489\\
};
\addplot [color=black,dashed,forget plot]
  table[row sep=crcr]{%
0.1	0.154023063187255\\
0.11	0.172346799030312\\
0.12	0.207860941569304\\
0.13	0.275838931181645\\
0.14	0.306245008066467\\
0.15	0.317213106447749\\
0.16	0.313029315457588\\
0.17	0.337059663691265\\
0.18	0.376486232998513\\
0.19	0.404597255309925\\
0.2	0.438593242425775\\
0.21	0.47279520957604\\
0.22	0.494685904638653\\
0.23	0.516929560808451\\
0.24	0.532345919661145\\
0.25	0.539565955470529\\
0.26	0.537565350657551\\
0.27	0.537950663914319\\
0.28	0.54101871052547\\
0.29	0.542960847424504\\
0.3	0.551051639928104\\
0.31	0.562105618300852\\
0.32	0.577712675517555\\
0.33	0.576509278114344\\
0.34	0.573438334616578\\
0.35	0.574179200207784\\
0.36	0.554995355436835\\
0.37	0.556740355615036\\
0.38	0.542044301394371\\
0.39	0.529382631815357\\
0.4	0.527086127376786\\
0.41	0.536355861500567\\
0.42	0.549286004316664\\
0.43	0.540984153045748\\
0.44	0.540518241815262\\
0.45	0.548687715478725\\
0.46	0.55569396010277\\
0.47	0.559748899645712\\
0.48	0.551244895000427\\
0.49	0.537876591974179\\
0.5	0.522218877535948\\
0.51	0.515807694096312\\
0.52	0.513002639149617\\
0.53	0.507661323958005\\
0.54	0.487703178484934\\
0.55	0.471570093691384\\
0.56	0.451958573946662\\
0.57	0.437885868918614\\
0.58	0.432237723464803\\
0.59	0.430387711264568\\
0.6	0.422218735924333\\
0.61	0.412031393038657\\
0.62	0.402253414307673\\
0.63	0.402799789229623\\
0.64	0.415111043602383\\
0.65	0.422557249734961\\
0.66	0.420359482556973\\
0.67	0.416716111533947\\
0.68	0.40023644763305\\
0.69	0.388441448313536\\
0.7	0.378067872467007\\
0.71	0.369118936826012\\
0.72	0.364799401572404\\
0.73	0.371557511440871\\
0.74	0.375301533290629\\
0.75	0.377078661566164\\
0.76	0.371326167150431\\
0.77	0.372002534025736\\
0.78	0.369738860972787\\
0.79	0.364379960219366\\
0.8	0.359573374317\\
0.81	0.347810899749298\\
0.82	0.344416438380063\\
0.83	0.338734974315285\\
0.84	0.32334411052411\\
0.85	0.301175020085228\\
0.86	0.288592316710571\\
0.87	0.277347009878215\\
0.88	0.261054273567537\\
0.89	0.248148857753407\\
0.9	0.230494223162692\\
0.91	0.218405424147419\\
0.92	0.220107398444169\\
0.93	0.225895247579606\\
0.94	0.225332924953774\\
0.95	0.225837730841351\\
0.96	0.223386064225784\\
0.97	0.219053332896275\\
0.98	0.203600953835306\\
0.99	0.199650891624445\\
1	0.168123190357644\\
};
\addplot [color=black,dashed,forget plot]
  table[row sep=crcr]{%
0.1	0.129248671887806\\
0.11	0.1453712718647\\
0.12	0.178767873741155\\
0.13	0.247379759751177\\
0.14	0.277322199374909\\
0.15	0.293763300874961\\
0.16	0.28402478680584\\
0.17	0.310357170188381\\
0.18	0.35073708380752\\
0.19	0.378268598045265\\
0.2	0.414543859001164\\
0.21	0.445216856767425\\
0.22	0.469739745147625\\
0.23	0.491283198002576\\
0.24	0.509846694395373\\
0.25	0.513546076752242\\
0.26	0.510480722432\\
0.27	0.513342401803254\\
0.28	0.515811103608592\\
0.29	0.516256235896967\\
0.3	0.526910983190547\\
0.31	0.536059463802981\\
0.32	0.550938999376163\\
0.33	0.550987267355395\\
0.34	0.546174305728987\\
0.35	0.548753502399716\\
0.36	0.527292744846273\\
0.37	0.530916187042406\\
0.38	0.515755122474316\\
0.39	0.501247664948702\\
0.4	0.498039939354717\\
0.41	0.512785410906986\\
0.42	0.521890461826579\\
0.43	0.514971099036494\\
0.44	0.51351461896043\\
0.45	0.522228338865471\\
0.46	0.530419484126116\\
0.47	0.534855064906622\\
0.48	0.526008975155146\\
0.49	0.51276677667521\\
0.5	0.498106871548515\\
0.51	0.492322164955934\\
0.52	0.485514296825473\\
0.53	0.482331931868215\\
0.54	0.461681353967266\\
0.55	0.445844181104944\\
0.56	0.422360099446643\\
0.57	0.409730022166797\\
0.58	0.406519894297409\\
0.59	0.404611268140759\\
0.6	0.3979483401136\\
0.61	0.385814064446552\\
0.62	0.376339518474303\\
0.63	0.37680069731419\\
0.64	0.387297184255158\\
0.65	0.395644257372272\\
0.66	0.394046775477551\\
0.67	0.389197840948963\\
0.68	0.375960044251726\\
0.69	0.362438114608032\\
0.7	0.348487563506061\\
0.71	0.34289292652612\\
0.72	0.337855044055609\\
0.73	0.348122290772379\\
0.74	0.348357829404828\\
0.75	0.350934014485953\\
0.76	0.34456980439391\\
0.77	0.344501657938114\\
0.78	0.343283908402229\\
0.79	0.341607872526766\\
0.8	0.335526161298668\\
0.81	0.325046669998678\\
0.82	0.318960004866016\\
0.83	0.313471321680085\\
0.84	0.297411773120054\\
0.85	0.274622270847087\\
0.86	0.263095890315334\\
0.87	0.251009693111197\\
0.88	0.230806892083252\\
0.89	0.224440340689326\\
0.9	0.20297722991363\\
0.91	0.196946290933796\\
0.92	0.194136574230798\\
0.93	0.199938073069129\\
0.94	0.199741017812401\\
0.95	0.196718847605224\\
0.96	0.197638313097863\\
0.97	0.193642903024019\\
0.98	0.176620975693296\\
0.99	0.17350898238699\\
1	0.142807979652137\\
};
\node[right, align=left, inner sep=0mm, font=\fontsize{10pt}{1em}\selectfont, text=black]
at (axis cs:0.85,0.5,0) {$(c)$};
\end{axis}
\end{tikzpicture}%
\caption{LSE: Estimation of the amplitude of a noisy sine wave quantized with uniform and non--uniform quantizers: (a) LSE estimator bias associated to the usage of a $10$--bit uniform quantizer; (b) INL of a $10$--bit non--uniform quantizer used to obtain data shown in (c); LSE estimator bias associated to the usage of the $10$--bit non--uniform quantizer {\color{black} with the estimated  $\pm 1 \sigma$ band, graphed using dashed lines }(c). Each sample is obtained by using $100$ records of data with $N=2000$ and $\sigma=0.3 \Delta$. \label{mot}}
 \end{figure} 
%Absolute value of the approximation error of summation (\ref{mean}) by the integral %(\ref{approxim}), $e(N,R)$ as a function of $N$, for various values of the number of records $R$. A single transition value is considered, $T_k=1$ with $\theta=1$. \label{abserr}}
%\end{figure}
The problem described in subsection \ref{ps} has been {\color{black}customarily} addressed by 
applying the LSE to the available data. However, the LSE is not proved to be optimal in the mean--square sense, when data are quantized:  estimates may be affected by bias or minimum estimation variance is not attained. A major difference occurs if the quantizer is uniform or is not {\color{black}uniform}, i.e. if transition levels are equally and uniformly spread over the signal input interval.
While reasonable performance is provided by LSE when the quantizer is uniform, when integral non--linearity (INL) affects {\color{black} it}, traditional estimators 
become appreciably {\em biased} {\color{black} \cite{CarboneSchoukens}}.
This means that, on the average, the difference between estimates and $\theta$
is no longer negligible.    
To appreciate this effect, consider Fig.~\ref{mot}, where 
the estimator bias associated to the estimation of $\theta$ using an ideally uniform (a) and non--uniform (c) quantizer having INL shown in plot (b), is graphed normalized to $\Delta$ as a function of $\theta$. 
A $10$--bit monotone
quantizer was assumed and $100$ records, each containing $2000$
samples of input quantized data were used to perform {\color{black} Monte Carlo}--based simulations. Zero--mean 
Gaussian noise having standard deviation equal to $0.3 \Delta$ was further assumed.
Fig.~\ref{mot}(a) shows that when threshold levels are uniformly spaced, the estimation {\color{black} bias} is negligible compared to $\Delta$. On the contrary when the INL shown in Fig.~\ref{mot}(b) affects the quantizer the bias is no longer negligible with respect to $\Delta$, as shown by data in Fig.~\ref{mot}(c). Since, in practice, INL almost always affects the quantizer behavior, it is of interest to propose new estimators for $\theta$.  

The bias associated to the behavior of the LSE
 can be explained by observing that the LSE does not include information about the position of the transition levels, since it only minimizes a figure--of--merit based on the sum of squared errors. Thus, improvements in estimation performance can be obtained by including also information about INL when processing data.

Recall that noise superimposed on the input signal may act as a {\em dither} 
signal, linearizing the behavior of the quantizer and thus rendering the LSE closer to optimality when assuming uniform quantizers. However, when the quantizer is not uniform, the linearization effect induced by  small--amplitude {\em dithering} is not effective, as in this case dithering
smooth{\color{black}s} the input--output characteristics {\color{black}only} locally and not over large input intervals \cite{Kollarbias}\cite{CarbonePetri}. Thus, only a large increase in the noise standard deviation would have 
positive effects on the estimator bias shown in Fig.~\ref{mot}(c), at the expense of 
a large estimator variance and of a higher risk of  overloading the ADC.

In the next section it is shown how to exploit information on the INL and thus on the position of the quantizer transition levels to remove estimator bias.    
Solutions differ whether 
%$\lambda$ is or is not a rational number:
\begin{itemize}
\item samples are collected in a small number of groups: this case was treated in \cite{CarboneSchoukensMoschitta};
 \item samples cover the phase space densely: 
%this case was treated in \cite{CarboneSchoukensMoschitta} when the %samples are collected in groups;
%\item $\lambda$ 
%is an irrational number: 
the simplified approach taken in \cite{CarboneSchoukensMoschitta} cannot be adopted and a new estimator is presented here that is based on the evaluation of statistical moments.
\end{itemize}

%The bias scales with the number of bits. Thus,   
%If the noise standard deviation increases the estimator bias progressively and uniformly vanishes, although 

%\begin{figure}[h]
%\begin{center}
%%\includegraphics[scale=0.45]{./../figbias.eps}
%\caption{Arithmetic mean estimator. Bias in the estimation of a DC value in Gaussian noise with $\sigma=0.25 \Delta$. \label{figbias}}
%\end{center}
%\end{figure}  

\section{A Mean Value--Based Estimator (MVBE)}
Observe that each output sample carries some 
information about the parameter to be estimated. 
In fact by taking into consideration that for a given time index $n$ and noise sample
$\eta_n$ the ADC output is determined by the input value being lower or 
larger than each transition level, for all transition levels we can define the indicator variables:
\be
z_{n,k} =
\left\{
\begin{array}{lll}
1 & \theta x_n + \eta_n > T_k  & n=0, \ldots, N-1 \\
 & &  k=0, \ldots, L-1\\
0 & \mbox{otherwise} &
\end{array}
\right.
\label{znk}
\ee
Thus, $z_{n,k}$ is a Bernoulli random variable with probability of success 
$p_{n,k} = 1-F(T_k-\theta x_n)$, where $F(\cdot)$ represents the noise
cumulative distribution function.
The summation of $z_{n,k}$ over the entire set of samples available over time provides,
\be
	Z_{N,k} = \sum_{n=0}^{N-1}z_{n,k} \label{cum}
\ee
that is  random variable taking values in $[0, N]$. Observe that $Z_{N,k}$ is not a binomial random variable, because the {\em success} probability  
varies from sample to sample, as 
$x_n$ in the event in (\ref{znk})  depends on the time index $n$.
Notice also that a single instance of (\ref{cum})
%$
%	\frac{Z_{N,k}}{N}
%$
is an estimator of the mean value of $Z_{N,k}$,
that can formally be written as: 
\begin{align}
\begin{split}
	E(Z_{N,k}) & =  
	%\frac{1}{N}
	\sum_{n=0}^{N-1}E(z_{n,k})  \\
	& =
	 %\frac{1}{N}
	 \sum_{n=0}^{N-1} \left[ 1-F(T_k-\theta \sin(2\pi \lambda n +{\color{black} \phi_0}) )\right] \\
	 & =
	 %\frac{1}{N}
	 \sum_{n=0}^{N-1} 
	 \left[ 1-F\left(T_k-\theta \sin\left(2\pi \left \langle \lambda n +\frac{{\color{black} \phi_0}}{2\pi} \right\rangle \right) \right)\right] 
	 \label{mean}
\end{split}
\end{align}
where $E(\cdot)$ is the expectation operator 
%$F(\cdot)$ is the noise cumulative distribution function 
and $\left \langle \cdot \right \rangle$ is the fractional part operator.
This expression relates the value of the unknown parameter $\theta$ to 
$E(Z_{N,k})$. 
If the coefficient of variation of 
$E(Z_{N,k})$ is not too large, by the law of large numbers
$E(Z_{N,k})$
can be estimated by 
a single instance of (\ref{cum})
%$\overline{Z}_{N,k} = \frac{Z_{N,k}}{N}$, 
and the inversion of 
(\ref{mean}) could provide a value of $\theta$ once an estimate of 
$E(Z_{N,k})$ is available and all other parameters are known. 
The numerical inversion of 
(\ref{mean}) becomes cumbersome when the number of samples increases significantly 
and requires knowledge of both $\lambda$ and ${\color{black} \phi_0}$, when $\sigma>0$. 
In the next subsection it will be shown how to remove both limitations and how to obtain a good approximation when $\sigma \simeq 0$.

\subsection{An Approximation of (\ref{mean})}
Both the inversion of (\ref{mean}) when $N$ becomes large and
the necessity of {\color{black} knowing}  $\lambda$ and 
${\color{black} \phi_0}$, {\color{black}result in a}  procedure {\color{black} that is}
difficult to be applied in practice.
While {\color{black} writing }a simple exact expression for the sum in (\ref{mean})
%$g(\theta)$ 
appears {\color{black} as a difficult task}, 
a good approximation can be found either by using the Euler--Maclaurin formula \cite{GrahamKnuthPatashnik} or the equidistribution theorem \cite{Petersen}.
While the Euler--Maclaurin formula still requires knowledge of $\lambda$ and ${\color{black} \phi_0}$,
the equidistribution theorem states that
for any function $g(\cdot)$, and coefficients
$a$ and $b$, with $a$ being {\em irrational},
%result in ergodic theory 
\cite{Petersen}:
\be
\lim_{N\rightarrow \infty} 
\frac{1}{N} 
\sum_{n=1}^{N}g\left(\langle an+b \rangle\right)
 =\int_{0}^{1}g(u)du
\ee
Thus, when $\lambda$ is irrational,
\begin{align}
\begin{split}
\lim_{N\rightarrow \infty}	
\frac{1}{N} 
E(Z_{N,k}) & = 
\int_0^1 
\left[ 1-F\left(T_k-\theta \sin\left( 2\pi u \right) \right)\right]du \\
&
\eqqcolon 
\overline{E}(Z_k)
\label{weil}
\end{split}
\end{align}
so that, for sufficiently large values of $N$,  
%the right member in
 $\overline{E}(Z_k)$
%(\ref{weil})
 can be considered 
an approximation of $\frac{E(Z_{N,k})}{N}$.
{\color{black} Moreover, }by defining $U$ as a uniform random variable in $[0,1]$, 
this term 
can also be written as
\begin{align}
\begin{split}
 \overline{E}(Z_k) = &	E\left(   \left[ 1-F\left(T_k-\theta \sin\left(2\pi U \right) \right)\right]\right) \\
	= & E\left(   1-F(T_k-\theta X) \right)  
\label{approxim}
\end{split}
\end{align}
where $X=\sin(2\pi U)$ is a transformed random variable with 
a PDF characterized by an arcsin distribution, whose expression is given by
\be
	\def\arraystretch{1.5}
	f_X(x) = 
	\left\{
		\begin{array}{ll}
		\dfrac{1}{\pi \sqrt{1-x^2}} & -1 < x < 1 \\
		0 & \mbox{otherwise}
		\end{array}
	\right.
	\label{arcsinpdf}
\ee
Thus, from (\ref{approxim}) we have:
\be
	\overline{E}(Z_k) = 
	\int_{-1}^{1} \frac{1}{\pi \sqrt{1-x^2}} \left[ 1-F(T_k-\theta x)\right] dx
	\label{inte}
\ee
When $N$ is sufficiently large{\color{black},}  
$\overline{E}(Z_k) \simeq \frac{E(Z_{N,k})}{N}$, as defined in (\ref{mean}) where $E(Z_{N,k})$ is estimated by a single instance of $Z_{N,k}$.
To verify this statement, consider the absolute error sequence defined as
\be
	e(N,R) = \left|
	\overline{E}(Z_{k})- \frac{1}{NR}\sum_{i=1}^R Z_{N,k,i} 
	\right|
\ee 
where $R$ represents the number of records, {\color{black} each containing} $N$ samples and $Z_{N,k,i}$ represents the value of $Z_{N,k}$ in the $i$--th record.
Expression $e(N,R)$ is plotted in Fig.~\ref{abserr} as a function of $N=10^3, \ldots, 120\cdot 10^3$ for $R=10^3, 5\cdot 10^4$, when assuming $T_k=1$ and $\theta=1$.
Gaussian noise with known variance is assumed so that $F(x) = \Phi\left( \frac{x}{\sigma}\right)$, where $\Phi(\cdot)$ is the cumulative distribution function of a standard Gaussian random variable.
It can be observed that for a given number of records $R$, by increasing $N$, overall 
lower values of the absolute error are attained.

\begin{figure}
%\scalebox{1}{
\input{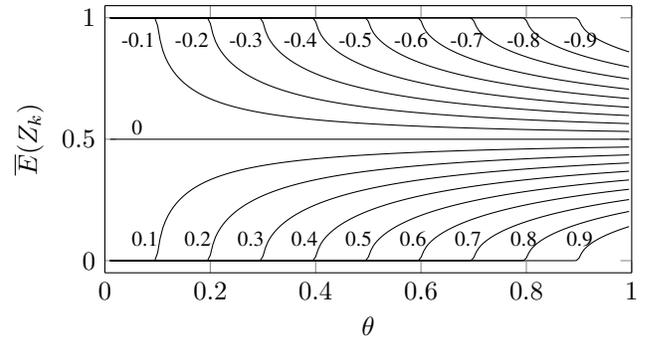}
\caption{Behavior of (\ref{inte}) as a function of $0.01 < \theta < 1$ for various values of
$-0.9 < T_k <0.9$. For a given value of $T_k$ and for a given estimate of 
$\overline{E}(Z_k)$, the corresponding curve is numerically inverted to provide a single estimate of $\theta$. \label{figfunction}}
\end{figure}

{\color{black}Observe that} $\overline{E}(Z_k)$ does not depend on ${\color{black} \phi_0}$, the initial record phase, which does not need to be known. 
For sufficiently large values of $N$,
$\frac{{Z}_{N,k}}{N}$ can be made equal to (\ref{inte}) so that the equality can be solved for $\theta$.  
To appreciate how the procedure operates, consider the behavior of (\ref{inte})
represented in Fig.~\ref{figfunction} as a function of $\theta$, for various values of
$-0.9 < T_k < 0.9$ and assuming zero--mean Gaussian noise with $\sigma=1.5\Delta$.
For a given value of $T_k$, the corresponding curve can be inverted to yield a value for
$\theta$ once a value on the $y$--axis is known. The calculation of $\overline{E}(Z_k)$
over all possible values of $k$ provides such information. 

Observe that curves cannot be inverted
in $3$ cases: when $\overline{E}(Z_k)=0, 0.5, 1$. Thus, the inversion procedures discards these values if they are returned by experiments.
%This is made clear in Fig.~\ref{fig function} by the straight line corresponding to $T_k=0$. 
%None of the other plotted curves intersect this line and there will not be values of $%\theta$ that result in (\ref{inte}) being equal to $0.5$.
%
A special case is the case $\overline{E}(Z_k)=0.5$.
When this occurs there are infinite solutions for $\theta$. 
This corresponds to the fact that 
(\ref{inte}) always provides the value $0.5$ when $T_k=0$, 
independently from the value of $\theta$.
Since derivatives of the curve{\color{black}s} in Fig.~\ref{figfunction} with respect to $\theta$
are close to $0$ when the mean value is close to $0.5$,
to maintain a safety margin that will guarantee possible numerical inversion of (\ref{inte}),
all values of $\overline{E}(Z_k)$ such that 
$|\overline{E}(Z_k)-0.5| <0.2$ will be discarded, where the threshold $0.2$ is determined heuristically. 
By iterating the inversion of (\ref{inte}) over all possible values of $T_k$
several estimate of $\theta$ results. The number of such estimates, defined in the 
following by $M$, equals the number of ADC transition levels, diminished by $1$ 
every time $\overline{E}(Z_k) = 0, 1$ or $|\overline{E}(Z_k)-0.5| <0.2$. 
The above procedure is true for each $T_k$. Thus, we have estimates of 
$\theta$ using each transition level. A straightforward combination of these estimates is their arithmetic mean; a better estimate can be derived by considering the variance of each singular estimate.
%Final estimate of $\theta$ is obtained by averaging the $M$ estimates.
The general estimation procedure is described 
by the pseudocode of Algorithm~\ref{algone} in the table.

As a final remark, consider that  
the solution is much simpler when 
there is no or very little noise {\color{black}that is when}  $\sigma \simeq 0$. In this case
$F(\cdot)$ can be approximated using a unity step function,
so that
(\ref{inte}) can be solved to yield:
\be
	\overline{E}(Z_k) \simeq -\frac{1}{\pi}
	\arcsin\left( \frac{T_k}{\theta}\right)
	+\frac{1}{2}, \quad \sigma \simeq 0
\label{appx}
\ee
By equating (\ref{appx}) to $\frac{Z_{N,k}}{N}$ and solving 
for $\theta$, when $0.2<\left|\frac{Z_{N,k}}{N}-0.5\right|<0.5$ we have:
\be
	\hat{\theta}_k \simeq 
	\frac{T_k}{\sin\left[ \left(\frac{1}{2}-\frac{Z_{N,k}}{N} \right)\pi\right]}, \quad \sigma \simeq 0
	\label{simple}
\ee
When $\sigma$ is not negligible, (\ref{appx}) is not accurate, since the arcsin function needs to be convolved by the noise PDF. In such cases, a good approximation can be obtained using a Taylor series expansion of the sin/arcsin function.
%{\color{black} What follows must be revised.
%Assume that $x$ has an arcsin pdf and define
%\be
%p=P(x<Tk) 
%\ee
%Define the count $c_i$ as the number of times  $x < Tk$ in $N$ experiments.
%Then the count $c_i$ is a binomial random variable with parameters
%$N$ and $p$. 
%Thus, $p_i=\nicefrac{c_i}{N}$ is a random variable with mean value $p$ and variance $p(1-p)/N$.
%If now we have $y=x+\eta$, $\eta$ Gaussian, then again define
%\be
%	p_1=P(y<Tk)
%\ee
%redefine $c_i$ as the count in this case 
%and a new $p_i = \nicefrac{c_i}{N}$ is a random variable with mean value $p_1$ and variance $p_1(1-p_1)/N$.
%Now define $z=(p_i-0.5)\pi$ and expand 
%\be
%	\sin((p_i-0.5)\pi) 
%\ee
%about the expected value of $z$, $z_0=E((p_i-0.5)\pi)=\pi(p_1-0.5)$
%\be
%	\sin((p_i-0.5)\pi) \simeq \sin(z_0)+\cos(z_0)(z-z_0)-0.5\sin(z_0)(z-z_0)^2
%\ee
%thus,
%\be
%	E\left( \sin((p_i-0.5)\pi) \right) \simeq \sin(z_0)+0-0.5\mbox{var}(z)
%\ee
%Now
%\be
%\mbox{var}(z) = \pi^2 \mbox{var}(p_i) = \pi^2 \frac{p_1(1-p_1)}{N}
%\ee
%}

\setlength\fheight{5.75cm} 
\setlength\fwidth{\fheight*\real{1.218}}

\begin{figure}
%\scalebox{1}{
% Minimal pgfplots version: 1.3
%  http://www.mathworks.com/matlabcentral/fileexchange/22022-matlab2tikz
\begin{tikzpicture}

\begin{axis}[%
width=\fwidth,
height=0.618034\fheight,
at={(0\fwidth,0\fheight)},
scale only axis,
separate axis lines,
every outer x axis line/.append style={black},
every x tick label/.append style={font=\color{black}},
xmin=1000,
xmax=120000,
xlabel={$N$},
every outer y axis line/.append style={black},
every y tick label/.append style={font=\color{black}},
ymode=log,
ymin=1e-08,
ymax=0.001,
yminorticks=true,
ylabel={$\left|e(N,R)\right|$}
]
\addplot [color=black,solid,line width=1.3pt,forget plot]
  table[row sep=crcr]{%
1000	4.1115188873508e-05\\
2000	7.1115188873511e-05\\
3000	5.1115188873511e-05\\
4000	0.000118634811126492\\
5000	0.000157315188873512\\
6000	4.4718144459825e-05\\
7000	4.0456239697916e-05\\
8000	7.009811126488e-06\\
9000	2.6559633317956e-05\\
10000	6.7115188873514e-05\\
11000	2.4024279782604e-05\\
12000	5.9448522206844e-05\\
13000	3.2115188873513e-05\\
14000	3.7456239697919e-05\\
15000	2.161814445982e-05\\
16000	9.3884811126488e-05\\
17000	5.652695357939e-05\\
18000	0.000115670744429065\\
19000	5.4358495337012e-05\\
20000	2.965188873515e-06\\
21000	5.3543760302081e-05\\
22000	4.1430265671945e-05\\
23000	1.1854319308294e-05\\
24000	2.03185554018e-06\\
25000	1.1004811126487e-05\\
26000	2.4730964972644e-05\\
27000	5.625551867228e-06\\
28000	6.063382555058e-06\\
29000	3.6195155954072e-05\\
30000	1.4518144459824e-05\\
31000	2.207391771647e-06\\
32000	1.3396438873513e-05\\
33000	2.0600037358362e-05\\
34000	2.5090693479432e-05\\
35000	8.34195398363e-06\\
36000	4.8059633317957e-05\\
37000	4.9385459143785e-05\\
38000	5.1911126915966e-05\\
39000	1.4064298305978e-05\\
40000	9.284811126491e-06\\
41000	1.9055542833803e-05\\
42000	6.2567569825894e-05\\
43000	8.280159963701e-06\\
44000	4.0706097964423e-05\\
45000	3.9062588904265e-05\\
46000	5.549971482208e-06\\
47000	5.3182683466914e-05\\
48000	8.2531855540177e-05\\
49000	1.2599096840774e-05\\
50000	2.3095188873508e-05\\
51000	5.3747556224525e-05\\
52000	4.1865580357257e-05\\
53000	1.7469716786865e-05\\
54000	2.3411485169808e-05\\
55000	3.1630265671941e-05\\
56000	3.8688382555059e-05\\
57000	1.9237995891055e-05\\
58000	4.1960016459716e-05\\
59000	1.2233832941307e-05\\
60000	9.34811126486e-07\\
61000	2.9048745552721e-05\\
62000	1.703454371222e-05\\
63000	8.535604777281e-06\\
64000	7.224563873511e-06\\
65000	6.8730964972639e-05\\
66000	1.3445417187098e-05\\
67000	2.9257945454844e-05\\
68000	1.3262247697043e-05\\
69000	2.3044231416344e-05\\
70000	5.970525412201e-06\\
71000	9.504529436349e-06\\
72000	2.1093144459819e-05\\
73000	3.1759024489951e-05\\
74000	1.0776703018384e-05\\
75000	2.0871477793158e-05\\
76000	4.9983609926141e-05\\
77000	2.2699604457929e-05\\
78000	1.406390682223e-05\\
79000	8.568355430283e-06\\
80000	2.1959811126487e-05\\
81000	9.633707392029e-06\\
82000	1.3494567224048e-05\\
83000	1.0137823174682e-05\\
84000	1.936617444936e-06\\
85000	5.4673046420609e-05\\
86000	3.035973917188e-06\\
87000	1.3253119907991e-05\\
88000	3.5910643418964e-05\\
89000	2.474739435307e-06\\
90000	9.295922237598e-06\\
91000	2.741953983629e-06\\
92000	3.7036985039533e-05\\
93000	1.4373253389641e-05\\
94000	7.051359086276e-06\\
95000	7.873083610353e-06\\
96000	1.573522206845e-06\\
97000	1.0506941450827e-05\\
98000	1.8384811126489e-05\\
99000	2.8177740419418e-05\\
100000	1.8435188873511e-05\\
101000	1.4907268081434e-05\\
102000	1.659320930408e-06\\
103000	2.8049859670178e-05\\
104000	1.6345958104277e-05\\
105000	1.819950778276e-06\\
106000	1.3605754911244e-05\\
107000	1.3735278416208e-05\\
108000	1.2689262947584e-05\\
109000	4.9205912043923e-05\\
110000	1.4048447490124e-05\\
111000	1.8992919234594e-05\\
112000	6.724096840771e-06\\
113000	1.3017843740769e-05\\
114000	1.6913434487544e-05\\
115000	7.163071996052e-06\\
116000	1.08949057526e-07\\
117000	1.5200658958985e-05\\
118000	2.4333963668861e-05\\
119000	6.36197276874e-07\\
};
\addplot [color=black,dashed,line width=1.3pt,forget plot]
  table[row sep=crcr]{%
1000	4.1804811126492e-05\\
2000	8.634811126486e-06\\
3000	1.0375188873514e-05\\
4000	2.2299811126489e-05\\
5000	1.3116811126486e-05\\
6000	5.578522206846e-06\\
7000	6.156239697915e-06\\
8000	2.345188873508e-06\\
9000	1.5427033348711e-05\\
10000	1.0447188873511e-05\\
11000	1.210791614624e-05\\
12000	3.048144459825e-06\\
13000	1.16749656582e-06\\
14000	9.089096840775e-06\\
15000	3.640522206846e-06\\
16000	7.016438873512e-06\\
17000	1.875399361785e-06\\
18000	6.068144459824e-06\\
19000	2.100452031403e-06\\
20000	6.948811126492e-06\\
21000	2.236141254462e-06\\
22000	1.169734328059e-06\\
23000	4.643941561272e-06\\
24000	3.516022206843e-06\\
25000	4.97611126492e-07\\
26000	4.6749656582e-07\\
27000	1.0667033348712e-05\\
28000	7.34811126488e-07\\
29000	2.008949057525e-06\\
30000	2.337477793157e-06\\
31000	2.233843384554e-06\\
32000	4.31686126491e-07\\
33000	4.037613115938e-06\\
34000	6.116365344103e-06\\
35000	1.148811126492e-06\\
36000	1.580188873508e-06\\
37000	2.837243558922e-06\\
38000	1.22112691596e-06\\
39000	2.672111950437e-06\\
40000	1.05018887351e-06\\
41000	1.701042532047e-06\\
42000	1.27192078869e-07\\
43000	1.030537710722e-06\\
44000	9.959811126489e-06\\
45000	8.580077762402e-06\\
46000	5.215623656123e-06\\
47000	1.097151552019e-06\\
48000	1.244772206845e-06\\
49000	6.71515404123e-07\\
50000	5.421188873513e-06\\
51000	6.5987597074e-08\\
52000	2.337503434181e-06\\
53000	2.497830382947e-06\\
54000	2.414818503139e-06\\
55000	1.768643418963e-06\\
56000	6.39831730653e-07\\
57000	2.257442705433e-06\\
58000	2.32775080405e-07\\
59000	1.823972097e-08\\
60000	6.70477793153e-07\\
61000	7.518253749436e-06\\
62000	8.27069191002e-07\\
63000	4.910426968749e-06\\
64000	2.23356112649e-06\\
65000	5.538881181204e-06\\
66000	4.98124947957e-06\\
67000	8.08094708579e-07\\
68000	1.4981112649e-07\\
69000	8.07652641627e-07\\
70000	5.565382555063e-06\\
71000	3.779977605908e-06\\
72000	4.81755570936e-07\\
73000	3.084777914608e-06\\
74000	2.856269954589e-06\\
75000	2.057611126489e-06\\
76000	1.524399399828e-06\\
77000	3.552591470911e-06\\
78000	1.145580357256e-06\\
79000	4.254429379839e-06\\
80000	2.227188873515e-06\\
81000	1.246045694389e-06\\
82000	7.418359605216e-06\\
83000	4.19285259054e-07\\
84000	6.8923649256e-07\\
85000	1.445071226451e-06\\
86000	5.55421431653e-07\\
87000	1.934499218342e-06\\
88000	8.69734328057e-07\\
89000	2.042114497273e-06\\
90000	2.124366682045e-06\\
91000	3.54041895717e-07\\
92000	1.461767648231e-06\\
93000	6.61425432653e-07\\
94000	4.285662190316e-06\\
95000	3.203547968594e-06\\
96000	2.532727793159e-06\\
97000	1.4982687944e-08\\
98000	2.095219289754e-06\\
99000	4.40441398762e-07\\
100000	1.777211126486e-06\\
101000	3.23703724998e-07\\
102000	2.80891436371e-06\\
103000	6.481790815256e-06\\
104000	7.20195741877e-07\\
105000	2.551379349701e-06\\
106000	3.78962458414e-07\\
107000	2.065843079116e-06\\
108000	1.26444075612e-06\\
109000	1.250499199884e-06\\
110000	1.011720217395e-06\\
111000	7.78504820186e-07\\
112000	3.687132555059e-06\\
113000	5.00368444873e-06\\
114000	6.23434487548e-07\\
115000	4.277623656117e-06\\
116000	6.082430252824e-06\\
117000	2.759682921363e-06\\
118000	8.78370448522e-07\\
119000	2.109853143292e-06\\
};
\node[right, align=left, inner sep=0mm, font=\fontsize{10pt}{1em}\selectfont, text=black]
at (axis cs:8000,0.0005,0) {$R= 10^3$};
\node[right, align=left, inner sep=0mm, font=\fontsize{10pt}{1em}\selectfont, text=black]
at (axis cs:8000,2e-07,0) {$R= 5\cdot 10^4$};
\end{axis}
\end{tikzpicture}%
\caption{Absolute value of the approximation error of summation (\ref{mean}) by the integral (\ref{approxim}), $e(N,R)$ as a function of $N$, for two values of the number of records $R$. A single transition value is considered, $T_k=1$ with $\theta=1$. \label{abserr}}
\end{figure}
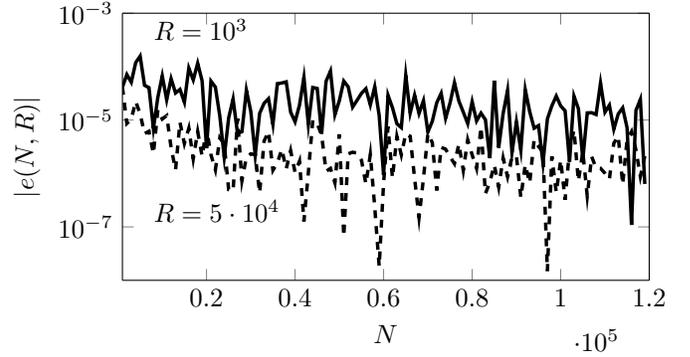

\begin{algorithm}
\caption{A procedure for the estimation of $\theta$ \label{algone}}
\label{theta}
\begin{algorithmic}[1]
\Procedure{Estimator}{$z_{n,k},T_k, \lambda, {\color{black} \phi_0}$}\Comment{Need all $T_k$'s}
 \For {$k \leftarrow 0, L-1$} \Comment{for every transition level}
 \For {$n \leftarrow 0, N-1$}  \Comment{for every sample} 
\State 		$z_{n,k} \leftarrow$ (count of samples $> T_k$)
		\EndFor
\State $\overline{Z}_{N,k}  \leftarrow \frac{1}{N} \sum_{n=0}^{N-1}z_{n,k}$ 
	\Comment{A count for every $k$}
\EndFor 
 % \State $r\gets a\bmod b$
 \State $M \leftarrow 0$ \Comment{now calculate several estimates of $\theta$ } 
\For {$k \leftarrow 0, L-1$} \Comment{for every count   $\overline{Z}_{N,k}$} 
    \If{$0 <\overline{Z}_{N,k}<1$ and $|\overline{Z}_{N,k}-0.5|>0.2$} 
    %\Comment{$0$, $1$ cannot be inverted}
%    \State \hskip-4mm $g(\theta) = 1-{\frac{1}{N}{\displaystyle \sum_{n=0}^{N-1}}}
%     F\left( T_k-\theta \sin\left( 2\pi \left \langle \lambda n +
   %  \frac{{\color{black} \phi_0}}{2\pi} \right\rangle \right) \right)$
    \State \hskip-4mm $g(\theta) = 
    \int_{-1}^{1} 
    \frac{1}{\pi \sqrt{1-x^2}} 
    \left[ 
    	1 -F(T_k-\theta x)
    \right] dx$
    \State \hskip-4mm $\hat{\theta}_M \leftarrow \theta$ 	such that
     %\right] 
     $g(\theta) =  \overline{Z}_{N,k}$ \Comment{one estimate}
     \State \hskip-4mm $M \leftarrow M+1$ \Comment{count the number of estimates}
   \EndIf\label{euclidendwhile}
   \EndFor
     \State $\hat{\theta} \leftarrow {\displaystyle 
     \frac{1}{M}\sum_{j=0}^{M-1} \hat{\theta}_j}$  \Comment{final estimate as the mean value} 
   \State \textbf{return} $\hat{\theta}$%\Comment{The gcd is b}
\EndProcedure
\end{algorithmic}
\end{algorithm}

\begin{figure}
%\scalebox{1}{
\input{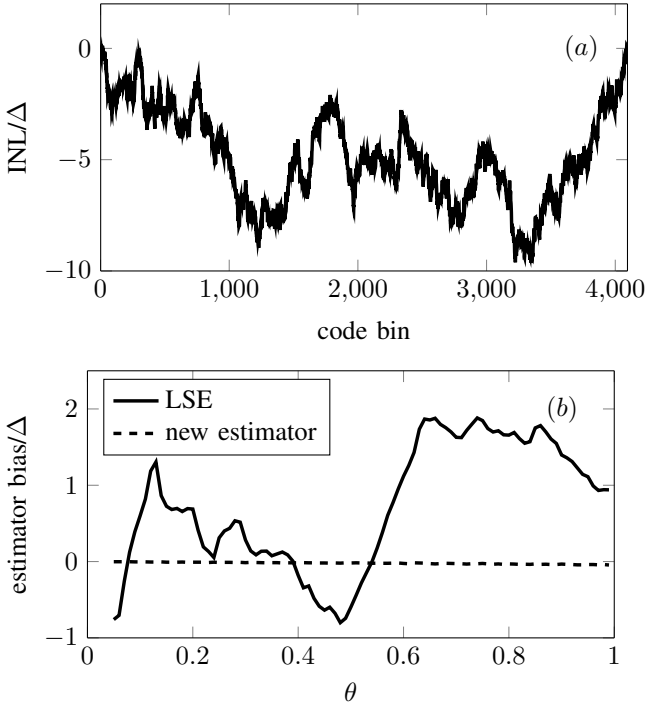}
\vskip3mm
% Minimal pgfplots version: 1.3
%  http://www.mathworks.com/matlabcentral/fileexchange/22022-matlab2tikz
\begin{tikzpicture}

\begin{axis}[%
width=\fwidth,
height=0.618034\fheight,
at={(0\fwidth,0\fheight)},
scale only axis,
separate axis lines,
every outer x axis line/.append style={black},
every x tick label/.append style={font=\color{black}},
xmin=0,
xmax=1,
xlabel={$\theta$},
every outer y axis line/.append style={black},
every y tick label/.append style={font=\color{black}},
ymin=-1,
ymax=2.5,
ylabel={estimator bias/$\Delta$},
legend style={at={(0.03,0.97)}, anchor=north west,legend cell align=left,align=left,draw=black,}
]
\addplot [color=black,solid,line width=1.3pt]
  table[row sep=crcr]{%
0.05	-0.759808000000007\\
0.06	-0.704511999999994\\
0.07	-0.256\\
0.08	0.122880000000009\\
0.09	0.391167999999993\\
0.1	0.600063999999975\\
0.11	0.819199999999995\\
0.12	1.18579200000002\\
0.13	1.304576\\
0.14	0.864255999999955\\
0.15	0.724991999999986\\
0.16	0.681984\\
0.17	0.696319999999957\\
0.18	0.65536000000003\\
0.19	0.694272000000012\\
0.2	0.688127999999949\\
0.21	0.411648000000014\\
0.22	0.192512000000022\\
0.23	0.131071999999961\\
0.24	0.0532479999999964\\
0.25	0.31129599999997\\
0.26	0.401407999999947\\
0.27	0.436223999999925\\
0.28	0.534527999999909\\
0.29	0.516096000000061\\
0.3	0.282624000000055\\
0.31	0.139264000000026\\
0.32	0.0880640000000312\\
0.33	0.135168000000021\\
0.34	0.137215999999967\\
0.35	0.0757760000000189\\
0.36	0.100352000000044\\
0.37	0.124927999999954\\
0.38	0.0880640000000312\\
0.39	0.0061439999999493\\
0.4	-0.180224000000067\\
0.41	-0.344064000000003\\
0.42	-0.319487999999978\\
0.43	-0.48127999999997\\
0.44	-0.585728000000017\\
0.45	-0.636928000000012\\
0.46	-0.59801600000003\\
0.47	-0.677887999999939\\
0.48	-0.800767999999948\\
0.49	-0.743424000000005\\
0.5	-0.59801600000003\\
0.51	-0.442367999999988\\
0.52	-0.28057600000011\\
0.53	-0.141311999999971\\
0.54	-0.00409600000011778\\
0.55	0.178175999999894\\
0.56	0.382975999999871\\
0.57	0.606208000000152\\
0.58	0.772096000000147\\
0.59	0.9523200000001\\
0.6	1.1182080000001\\
0.61	1.25952000000007\\
0.62	1.42950399999995\\
0.63	1.73465600000009\\
0.64	1.86777600000005\\
0.65	1.85548799999992\\
0.66	1.87801599999989\\
0.67	1.79814399999987\\
0.68	1.75513599999999\\
0.69	1.69984000000022\\
0.7	1.62815999999998\\
0.71	1.62406400000009\\
0.72	1.72032000000013\\
0.73	1.80223999999998\\
0.74	1.88211200000001\\
0.75	1.84729599999991\\
0.76	1.74284800000009\\
0.77	1.69779199999994\\
0.78	1.71417599999995\\
0.79	1.66092800000001\\
0.8	1.65683199999989\\
0.81	1.69164799999999\\
0.82	1.61177600000019\\
0.83	1.55033600000002\\
0.84	1.57081600000015\\
0.85	1.7510400000001\\
0.86	1.78176000000008\\
0.87	1.69983999999999\\
0.88	1.60563200000001\\
0.89	1.54828799999996\\
0.9	1.39673599999992\\
0.91	1.359872\\
0.92	1.31071999999995\\
0.93	1.22675199999981\\
0.94	1.14278400000012\\
0.95	1.11206400000015\\
0.96	0.991232000000082\\
0.97	0.931839999999966\\
0.98	0.942080000000033\\
0.99	0.940031999999974\\
};
\addlegendentry{LSE};

\addplot [color=black,dashed,line width=1.3pt]
  table[row sep=crcr]{%
0.05	-0.00204800000000205\\
0.06	-0.00204800000000205\\
0.07	-0.00204800000000205\\
0.08	-0.00204800000000205\\
0.09	-0.0040960000000041\\
0.1	-0.0040960000000041\\
0.11	-0.0040960000000041\\
0.12	-0.00614399999997772\\
0.13	-0.00614400000000614\\
0.14	-0.0040960000000041\\
0.15	-0.00819200000000819\\
0.16	-0.00819200000000819\\
0.17	-0.00819200000000819\\
0.18	-0.00614400000000614\\
0.19	-0.0102400000000102\\
0.2	-0.00819200000000819\\
0.21	-0.00819200000000819\\
0.22	-0.00819200000000819\\
0.23	-0.0122880000000123\\
0.24	-0.00819200000000819\\
0.25	-0.0122880000000123\\
0.26	-0.0102400000000671\\
0.27	-0.0102400000000671\\
0.28	-0.0122880000000123\\
0.29	-0.0102399999999534\\
0.3	-0.0143359999999575\\
0.31	-0.0102399999999534\\
0.32	-0.0122880000000123\\
0.33	-0.0143360000000712\\
0.34	-0.0163840000000164\\
0.35	-0.0143359999999575\\
0.36	-0.0143359999999575\\
0.37	-0.0184319999999616\\
0.38	-0.0163840000000164\\
0.39	-0.0163840000000164\\
0.4	-0.0163840000000164\\
0.41	-0.0143359999999575\\
0.42	-0.0163840000000164\\
0.43	-0.0163840000000164\\
0.44	-0.0184319999999616\\
0.45	-0.0184320000000753\\
0.46	-0.0204800000000205\\
0.47	-0.0184319999999616\\
0.48	-0.0204799999999068\\
0.49	-0.0163840000000164\\
0.5	-0.0184319999999616\\
0.51	-0.0163840000000164\\
0.52	-0.0184320000000753\\
0.53	-0.0184320000000753\\
0.54	-0.0225279999999657\\
0.55	-0.0204800000001342\\
0.56	-0.0225280000001931\\
0.57	-0.0225279999999657\\
0.58	-0.0204799999999068\\
0.59	-0.0184319999998479\\
0.6	-0.0245760000000246\\
0.61	-0.028671999999915\\
0.62	-0.0163840000000164\\
0.63	-0.0184320000000753\\
0.64	-0.0204800000001342\\
0.65	-0.0266240000000835\\
0.66	-0.0307199999999739\\
0.67	-0.0266240000000835\\
0.68	-0.0225280000001931\\
0.69	-0.0245759999997972\\
0.7	-0.0266239999998561\\
0.71	-0.028671999999915\\
0.72	-0.0327680000000328\\
0.73	-0.028671999999915\\
0.74	-0.028671999999915\\
0.75	-0.0245760000000246\\
0.76	-0.0307199999999739\\
0.77	-0.0348160000000917\\
0.78	-0.0307199999999739\\
0.79	-0.0368640000001506\\
0.8	-0.0307200000002013\\
0.81	-0.0327680000000328\\
0.82	-0.0368639999999232\\
0.83	-0.0348159999998643\\
0.84	-0.0368639999999232\\
0.85	-0.0307199999999739\\
0.86	-0.0368639999999232\\
0.87	-0.0348160000000917\\
0.88	-0.0389119999999821\\
0.89	-0.0389119999999821\\
0.9	-0.0327680000000328\\
0.91	-0.0368640000001506\\
0.92	-0.0389119999999821\\
0.93	-0.0430080000000999\\
0.94	-0.0430079999998725\\
0.95	-0.0389119999999821\\
0.96	-0.0389119999999821\\
0.97	-0.0430079999998725\\
0.98	-0.0389119999999821\\
0.99	-0.0450559999999314\\
};
\addlegendentry{new estimator};

\node[right, align=left, inner sep=0mm, font=\fontsize{10pt}{1em}\selectfont, text=black]
at (axis cs:0.87,2,0) {$(b)$};
\end{axis}
\end{tikzpicture}%
  %firgresult
%\hskip-1.6cm
\caption{Simulation results obtained with a monotonous $12$--bit non uniform ADC: (a)  
INL normalized to $\Delta$ as a function of the code--bin; 
(b)  normalized absolute estimation error as a function of the sine wave amplitude in the case of the LSE and of the MVBE.
\label{simINL}
}
\end{figure}

\section{Estimator Properties}
The properties of the MVBE, {\color{black} as resulting from the solution of (\ref{inte}), } were 
determined both by simulations and measurements.
\subsection{Simulation Results}
%The properties of the proposed estimator are first determined using %simulation results.
Algorithm~\ref{algone} was first implemented in $C$--code using the GNU Scientific Library that was needed for the numerical calculation of the integral in (\ref{inte}) and for its inversion. 
%Matlab and applied on data obtained from an $8$ 
A $12$--bit non--uniform ADC was simulated 
by using a resistor ladder characterized by normally distributed resistance values
to realize the $2^{12}-1$ transition levels. 
This approach guarantees monotonicity of the simulated ADC and allows values of
the INL  greater than $\Delta$. The behavior of INL normalized to $\Delta$, 
is plotted in Fig.~\ref{simINL}(a), as a function of the code bin.
{\color{black} Sine waves} with 
amplitude $\theta$ 
varying between $0.05$ and $1$ 
{\color{black} were} assumed as the input signal,  
and $R=10$ records of $N=32193$ samples each were collected and processed
with $\lambda=1050\pi/N$ and $\sigma=0.21\Delta$.
Results obtained by using the LSE and the MVBE are
shown in Fig.~\ref{simINL}(b), 
where the estimator bias 
normalized to $\Delta$, is plotted for  both cases
using a solid and a dashed line, respectively. 
It can be observed that the MVBE removes 
the bias associated to the behavior of the LSE.
In this case, the additional error associated to the usage of the simplified version of this estimator provided by (\ref{simple}) is negligible for all practical purposes with the exclusion of very small values of $\theta$.
In this latter case the number of excited thresholds is limited and neglecting the effect of noise produces a small detectable 
difference  between the two estimation approaches.

\begin{figure}
%\scalebox{1}{
% Minimal pgfplots version: 1.3
%  http://www.mathworks.com/matlabcentral/fileexchange/22022-matlab2tikz
\begin{tikzpicture}

\begin{axis}[%
width=\fwidth,
height=0.618034\fheight,
at={(0\fwidth,0\fheight)},
scale only axis,
separate axis lines,
every outer x axis line/.append style={black},
every x tick label/.append style={font=\color{black}},
xmin=0.24,
xmax=0.51,
xlabel={$\theta$},
every outer y axis line/.append style={black},
every y tick label/.append style={font=\color{black}},
ymin=-0.0942560000000001,
ymax=0.158276,
ylabel={estimator bias/$\Delta$},
legend style={at={(0.97,0.03)},anchor=south east,legend cell align=left,align=left,draw=black}
]
\addplot [color=black,solid,line width=1.3pt]
  table[row sep=crcr]{%
0.27	-0.0942560000000001\\
0.28	-0.0595800000000001\\
0.29	-0.0255299999999999\\
0.3	0.00674399999999997\\
0.31	0.036632\\
0.32	0.063774\\
0.33	0.087652\\
0.34	0.107372\\
0.35	0.12409\\
0.36	0.137166\\
0.37	0.146824\\
0.38	0.153012\\
0.39	0.156814\\
0.4	0.158276\\
0.41	0.157392\\
0.42	0.154552\\
0.43	0.150142\\
0.44	0.14419\\
0.45	0.137326\\
0.46	0.129118\\
0.47	0.120098\\
};
\addlegendentry{LSE};

\addplot [color=black,dashed,line width=1.3pt]
  table[row sep=crcr]{%
0.27	1.99999999994649e-06\\
0.28	4.7999999999937e-05\\
0.29	4.59999999999905e-05\\
0.3	3.79999999999825e-05\\
0.31	-3.39999999999785e-05\\
0.32	2.99999999999745e-05\\
0.33	0.000107999999999997\\
0.34	-0.000138000000000083\\
0.35	1.80000000000735e-05\\
0.36	0.000132000000000021\\
0.37	0.000174000000000007\\
0.38	-0.000129999999999963\\
0.39	-0.000103999999999993\\
0.4	4.7999999999937e-05\\
0.41	1.80000000000735e-05\\
0.42	-5.1999999999941e-05\\
0.43	-4.99999999999945e-05\\
0.44	-0.000210000000000043\\
0.45	4.99999999999945e-05\\
0.46	4.3999999999933e-05\\
0.47	0.000280000000000058\\
};
\addlegendentry{new estimator};

\end{axis}
\end{tikzpicture}%
%
%\input{./figure/figresult.tex}
%\hskip-1.6cm
\caption{Simulation results obtained with a monotonous $2$--bit ideally uniform ADC: normalized estimator bias  as a function of the sine wave amplitude in the case of the LSE and of MVBE ($\sigma=0.12\Delta$, $\lambda=0.723457$, ${\color{black} \phi_0}=0.4876$, $N=106777$).
\label{sim2bit}
}
\end{figure}
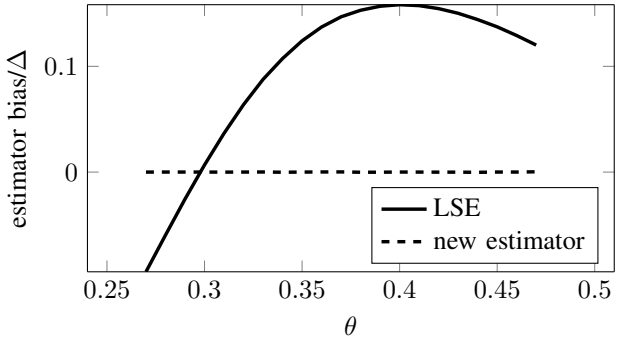

Consider that, being based on the knowledge of the threshold levels, the MVBE is characterized by a negligible 
bias even if severe quantization is performed. To prove this statement a $2$--bit uniform quantizer was assumed and the algorithm was applied with parameters: $\sigma=0.12\Delta$, $\lambda=0.723457$, ${\color{black} \phi_0}=0.4876$, $N=106777$.  
The estimator bias in the case of the MVBE and the LSE is shown in Fig.~\ref{sim2bit}. The LSE not being optimal in this case performs very poorly while the MVBE provides a very good performance.

\begin{figure}
%\scalebox{1}{
\input{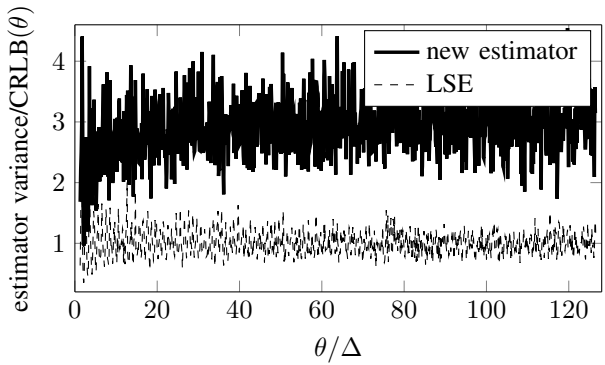}
%\hskip-1.6cm
%}
\caption{Simulation results obtained with an $8$--bit ideally uniform ADC: variance of the MVBE and of the LSE normalized to the corresponding CRLB
as a function of $\theta/\Delta$ ($\sigma=0.2\Delta$, $\lambda=0.1234$, $N=1000$).
 \label{antonio}}
 \end{figure}

{\color{black}
Finally, the variances of MVBE and LSE were 
evaluated against the Cramer--Rao lower bound {\color{black} (CRLB)}, by assuming an $8$--bit 
uniform ADC. The CRLB was calculated by the same approach described in \cite{CarboneMoschitta}, without resorting to 
the simplifying assumption introduced by
noise model of quantization \cite{KollarBook}.
Variances were normalized to the corresponding CRLB as a function of $\theta/\Delta$. Results based on $100$ records
obtained with  $\sigma=0.2\Delta$ and $N=1024$ are shown in Fig.~\ref{antonio}. Simulations show that because of its bias, the variance of 
LSE becomes smaller than the CRLB for some values of $\theta/\Delta$.
Conversely, MVBE is capable to reduce the bias at the expense of a larger than the CRLB variance.
%the bias removal properties of MVBE and 
%make it an estimator with a larger variance. However, the reduction in %statistical efficiency is more than acceptable in the considered case.
}

\begin{figure}
\begin{center}
\includegraphics[scale=0.45]{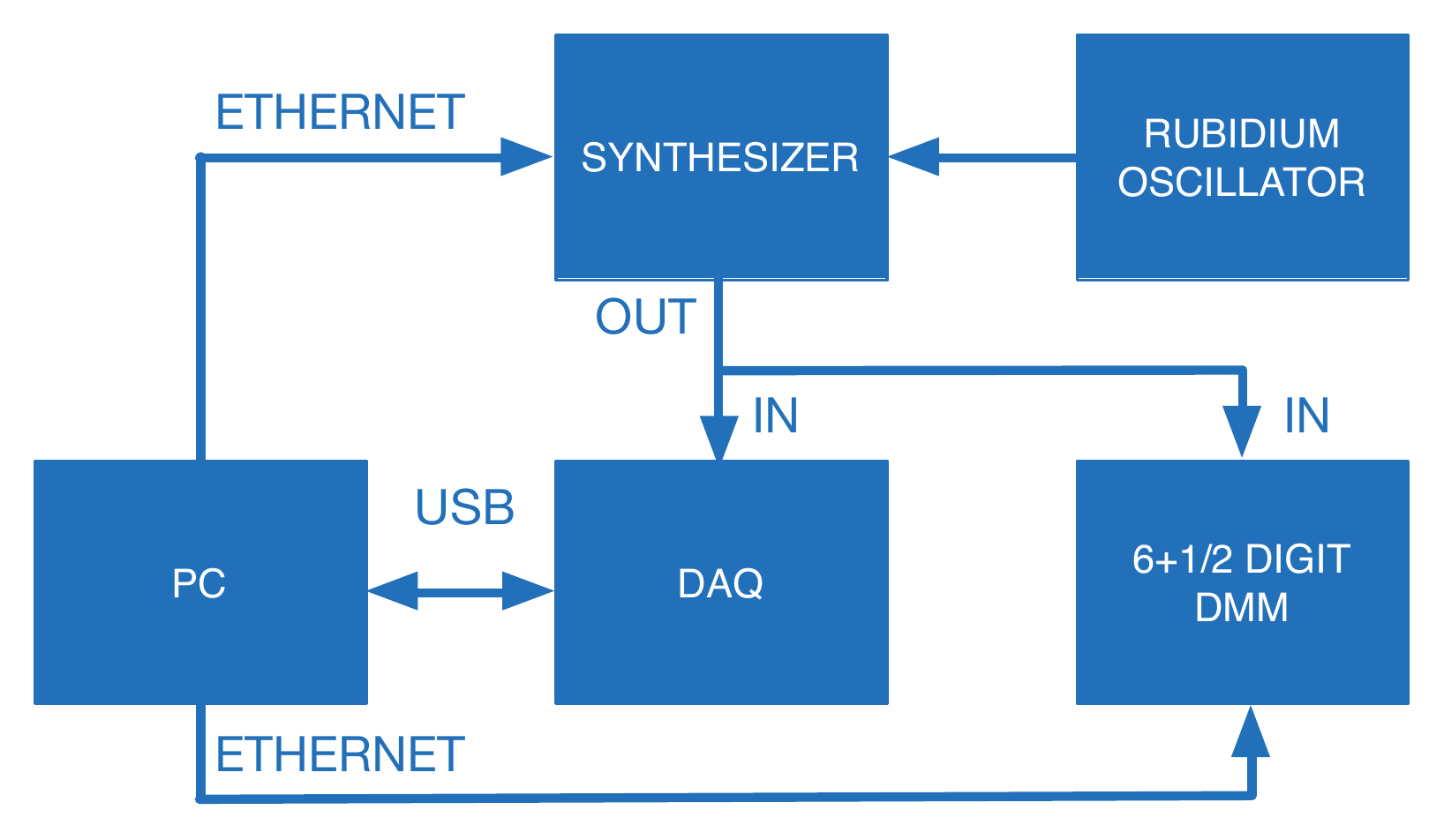}
\caption{The measurement set--up used to obtain experimental data.
The controlled synthesizer sources a sine wave to the 
$12$--bit data acquisition board, whose amplitude is measured by the digital multimeter in AC mode. A PC uses Ethernet and USB to control the measurement chain.
\label{expres}}
\end{center}
\end{figure}

\subsection{Experimental Results}
\begin{figure}
\begin{center}
\includegraphics[scale=0.45]{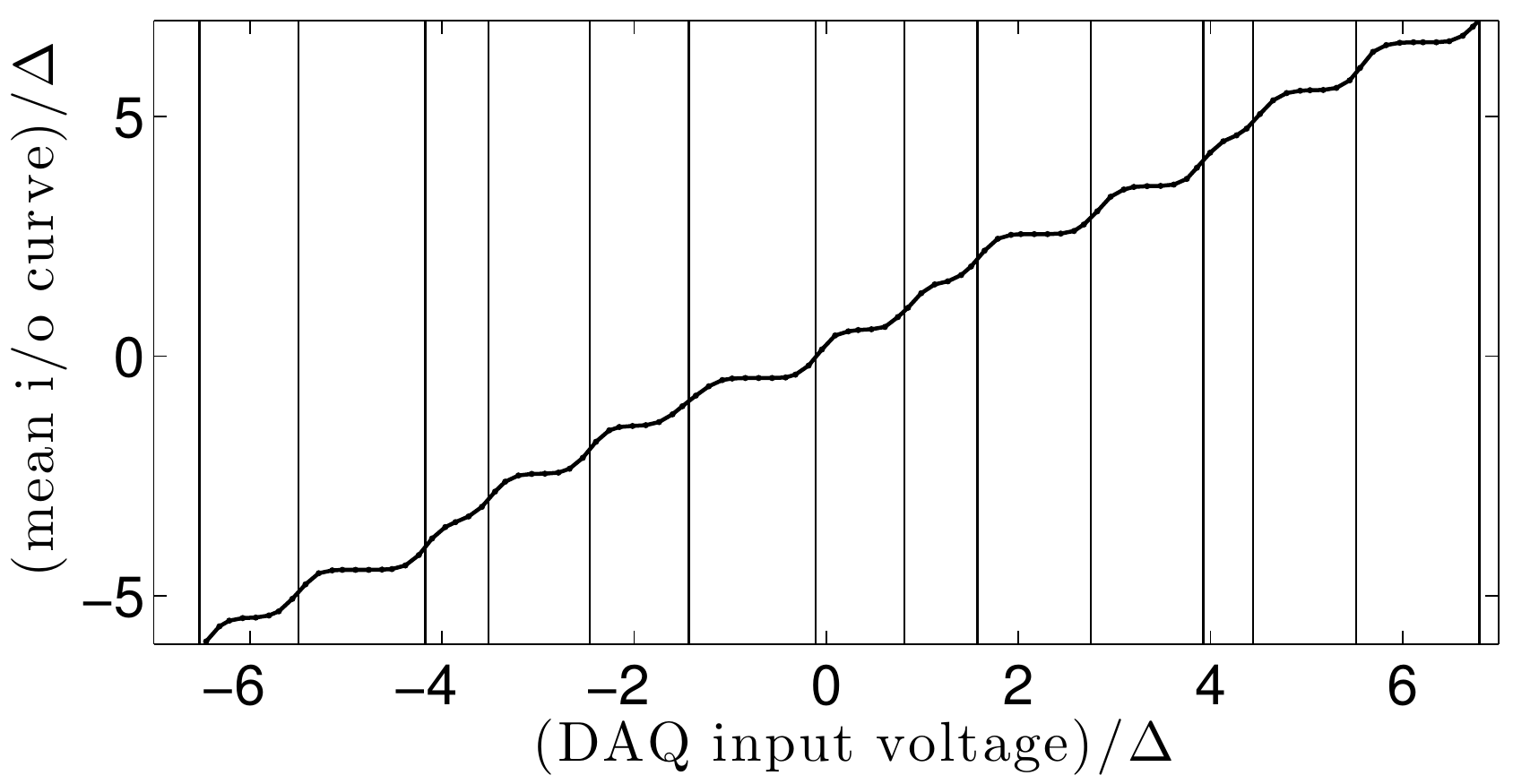}
\caption{Normalized mean value of the DAQ input/output characteristic as a function of the normalized input voltage (bold solid line). Shown are also estimated transition voltages normalized to $\Delta$ (thin vertical solid lines).
Experimental results are obtained by averaging $2000$ records, each based on $1200$ DC input voltage values provided by the synthesizer shown in Fig.~\ref{expres}.  The input was measured by the DMM in averaging mode, to provide the reference value shown on the $x$--axis.
\label{exp_io}}
\end{center}
\end{figure}

To prove the validity of the MVBE proposed in this paper, experimental results were obtained using the measurement chain depicted in Fig.~\ref{expres}. A rubidium source {\color{black} (Standford Research Systems PRS10)} controlled 
the waveform synthesizer {\color{black}(Agilent 33220A)} used to generate the sine wave signal 
fed to a $12$--bit commercial data acquisition board (DAQ, {\color{black} National Instruments NI6008)}. The instruments were connected to a portable PC using the Ethernet network, while the DAQ board was connected using the Universal Serial Bus (USB).
A {\color{black} $6\nicefrac{1}{2}$ digit} multimeter (DMM, {\color{black} Keithley 8845A}) was used as a reference instrument to obtain 
an accurate value of the generated signal amplitude, {\color{black} given that its accuracy is in the order of $0.06\%$ of the measurement result in the used measurement range ($100$ mV)}.
This setup was first used to estimate the 
transition levels of the ADC embedded in the DAQ and its voltage gain.
The values of the transition levels, normalized to $\Delta=5.096$ mV are shown in Fig.~\ref{exp_io} using thin solid  vertical lines. In the same figure {\color{black} it is} shown the mean input/output characteristics of the DAQ, obtained by averaging $2000$ records of $1200$ input voltage values, distributed in the interval $[-0.3986, 0.4007]$ V. The shape of the input/output curve 
does not show the typical staircase behavior associated to a perfect quantizer because of the smoothing effect of wide--band noise as 
shown in \cite{Kollarbias,Hejn}. Observe also that the transition voltages are not uniformly spaced, 
thus causing 
nonlinearity in the quantizer.

Also the standard deviation of the equivalent input--referred
noise source  $\sigma$ was determined as being about equal to $0.17\Delta$.
It must be observed that a much larger value of $\sigma$ resulted
when the {\color{black} screen of the } portable PC was used, because of electromagnetic disturbances. The best measurement condition was obtained by using an external monitor. 
%The INL and DNL of the DAQ were estimated to be 

This setup was then used to collect $3$ records of $N=287431$ samples of a sine wave with frequency $99.3715$ Hz, sampled at a nominal sampling rate equal to $9135$ samples per second, so that $\lambda = 0.0108781\ldots$ resulted. Records were taken by varying $\theta/\Delta$ in the interval $(21,52)$. In this interval, the magnitudes of the measured INL and DNL of the DAQ were all upper bounded by $\Delta/2$, thus making the ADC very linear.

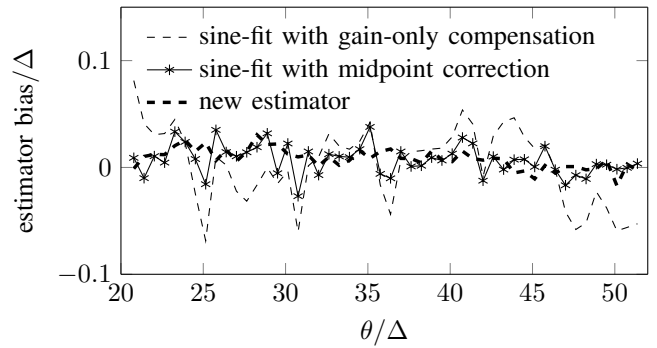
\begin{figure}[t]
%\scalebox{1}{
%\begin{rotate}{90}
%\includegraphics[scale=0.3]{../NI6008/thisone.eps}
% Minimal pgfplots version: 1.3
%  http://www.mathworks.com/matlabcentral/fileexchange/22022-matlab2tikz
\begin{tikzpicture}

\begin{axis}[%
width=\fwidth,
height=0.618034\fheight,
at={(0\fwidth,0\fheight)},
scale only axis,
separate axis lines,
every outer x axis line/.append style={black},
every x tick label/.append style={font=\color{black}},
xmin=20,
xmax=52,
xlabel={$\theta/\Delta$},
every outer y axis line/.append style={black},
every y tick label/.append style={font=\color{black}},
ymin=-0.1,
ymax=0.15,
ylabel={$\mbox{estimator bias}/\Delta$},
legend style={at={(0.03,0.97)},anchor=north west,legend cell align=left,align=left,fill=none,draw=none}
]
\addplot [color=black,dashed]
  table[row sep=crcr]{%
20.7815934065934	0.0813058205372077\\
21.4042386185243	0.0410784002331833\\
22.0281920460492	0.0307077020219401\\
22.651949241235	0.0317210978982397\\
23.2746598639456	0.0449669356498032\\
23.8988749345892	0.0234325902792337\\
24.5219126111983	-0.024796889642578\\
25.1469126111983	-0.0686721177548893\\
25.769819466248	0.0100013863130707\\
26.3941653584511	0.00219102989674981\\
27.0168759811617	-0.0222576753089596\\
27.6413526949241	-0.0315920185224831\\
28.2643903715332	-0.0164844560725408\\
28.8891287284144	-0.000275083753050122\\
29.5128859236002	-0.0150879320281921\\
30.1372972265829	-0.00338756469909314\\
30.7600078492936	-0.0603043096270508\\
31.3846807953951	0.00152171872831719\\
32.0075222396651	0.0099098902986384\\
32.632326007326	0.0314658112143189\\
33.2550366300366	0.0200301243344357\\
33.8802328623757	0.0170730975810423\\
34.5035975928833	0.0244758717382712\\
35.1282705389848	0.0421130975047888\\
35.7516352694924	-0.0163839007722488\\
36.3765698587127	-0.043824453161069\\
37.0001962323391	0.00679931131980853\\
37.6236917844061	0.0147005719100352\\
38.2486917844061	0.0156797962653779\\
38.8724489795918	0.0171831296145335\\
39.4977760334903	0.0180792998492461\\
40.1211407639979	0.0286218738103257\\
40.7458791208791	0.0537951090943785\\
41.3697017268446	0.0406714208693814\\
41.9947017268446	-0.00849335073363165\\
42.6185243328101	0.0294260057953098\\
43.2443092621664	0.0426972430712865\\
43.8831109366824	0.0465122469647705\\
44.5071297749869	0.0285423334708965\\
45.1337650444793	0.0176969529185331\\
45.7566064887493	0.0211150934371821\\
46.3825876504448	-0.00207645122845549\\
47.0045787545788	-0.041540793761568\\
47.6310177917321	-0.0580688932113298\\
48.2532705389848	-0.0518648227269279\\
48.8790554683412	-0.022738672431524\\
49.5	-0.0378442636746892\\
50.126504447933	-0.0593060601154939\\
50.7481684981685	-0.0561366908226282\\
51.3742804814233	-0.0527374770617905\\
};
\addlegendentry{sine-fit with gain-only compensation};

\addplot [color=black,solid,mark=asterisk,mark options={solid}]
  table[row sep=crcr]{%
20.7815934065934	0.00900524404327071\\
21.4042386185243	-0.0100055034951079\\
22.0281920460492	0.0105779331944725\\
22.651949241235	0.00450996737824834\\
23.2746598639456	0.0335260440206428\\
23.8988749345892	0.0238766279283871\\
24.5219126111983	0.00789334031720005\\
25.1469126111983	-0.015680770999758\\
25.769819466248	0.0351516960958643\\
26.3941653584511	0.0144144182453706\\
27.0168759811617	0.0104651327024711\\
27.6413526949241	0.0141885851601433\\
28.2643903715332	0.0189476200636646\\
28.8891287284144	0.0319374916381429\\
29.5128859236002	-0.00541814727056708\\
30.1372972265829	0.022394727198716\\
30.7600078492936	-0.0268855211229529\\
31.3846807953951	0.0149653122619541\\
32.0075222396651	-0.00713705395060343\\
32.632326007326	0.0124110217476037\\
33.2550366300366	0.0103222895048743\\
33.8802328623757	0.00961929657119662\\
34.5035975928833	0.0171227584817447\\
35.1282705389848	0.0379215595581274\\
35.7516352694924	-0.00601294990640695\\
36.3765698587127	-0.0101630408100028\\
37.0001962323391	0.0149347869386753\\
37.6236917844061	0.00095584969107533\\
38.2486917844061	0.00159604686039086\\
38.8724489795918	0.00922235442220005\\
39.4977760334903	0.00651457249007912\\
40.1211407639979	0.0129046893946035\\
40.7458791208791	0.0280194177336251\\
41.3697017268446	0.0225142929005031\\
41.9947017268446	-0.0123965436632159\\
42.6185243328101	0.0101526893618134\\
43.2443092621664	-0.00196713804185318\\
43.8831109366824	0.0073004812240618\\
44.5071297749869	0.00761369744941381\\
45.1337650444793	0.000512266596118416\\
45.7566064887493	0.0199212277801197\\
46.3825876504448	-0.00209924503612769\\
47.0045787545788	-0.0166834836077995\\
47.6310177917321	-0.00719167321510451\\
48.2532705389848	-0.0107343108158437\\
48.8790554683412	0.00311181026594193\\
49.5	0.00254423929361831\\
50.126504447933	-0.00162740100878064\\
50.7481684981685	-0.0003566187090856\\
51.3742804814233	0.00361044802950746\\
};
\addlegendentry{sine-fit with midpoint correction};

\addplot [color=black,dashed,line width=1.3pt]
  table[row sep=crcr]{%
20.7815934065934	-0.00107248713308703\\
21.4042386185243	0.0102798016067757\\
22.0281920460492	0.0121945512159601\\
22.651949241235	0.0119909090819051\\
23.2746598639456	0.020374724729538\\
23.8988749345892	0.0255380672172752\\
24.5219126111983	0.0150851971703278\\
25.1469126111983	0.023284319773238\\
25.769819466248	0.00643312712575565\\
26.3941653584511	0.0164960996447924\\
27.0168759811617	0.00580782800565142\\
27.6413526949241	0.0147911164384973\\
28.2643903715332	0.0305050241272299\\
28.8891287284144	0.021517632575849\\
29.5128859236002	0.0219693291332923\\
30.1372972265829	0.0124786699883398\\
30.7600078492936	0.00987225668533616\\
31.3846807953951	0.0115639037630154\\
32.0075222396651	0.00140648808108326\\
32.632326007326	0.00983021698751689\\
33.2550366300366	0.00203823904742177\\
33.8802328623757	0.00661358830049454\\
34.5035975928833	0.0157285869560407\\
35.1282705389848	0.00854296107139664\\
35.7516352694924	0.0148823162554101\\
36.3765698587127	0.017266837772097\\
37.0001962323391	0.00847395154971823\\
37.6236917844061	0.00833683224559926\\
38.2486917844061	0.00434743171263362\\
38.8724489795918	0.015352225839745\\
39.4977760334903	0.00725773261942818\\
40.1211407639979	0.00544448713706858\\
40.7458791208791	0.0152919092495218\\
41.3697017268446	0.00834999159154488\\
41.9947017268446	0.0068219494272054\\
42.6185243328101	0.00868015561419639\\
43.2443092621664	0.00815727887759722\\
43.8831109366824	-0.00513377586525158\\
44.5071297749869	-0.00329934240291007\\
45.1337650444793	-0.0106995940338878\\
45.7566064887493	0.00182296731502534\\
46.3825876504448	-0.00526180828459592\\
47.0045787545788	0.000810064301228385\\
47.6310177917321	0.000801846700570355\\
48.2532705389848	-0.00222526798718908\\
48.8790554683412	-0.00116327899963332\\
49.5	0.00676863779951538\\
50.126504447933	-0.0170341990570466\\
50.7481684981685	0.0030121022150929\\
51.3742804814233	-0.00272654289715373\\
};
\addlegendentry{new estimator};

\end{axis}
\end{tikzpicture}%
%\end{rotate}
\caption{Experimental results based on a $12$--bit commercial data acquisition board: estimator bias as a function of the sine wave amplitude $\theta$, both normalized to a nominal value $\Delta = 5.096$ mV.
Comparison between the performances of the MVBE and of $4$--parameter sine--fit estimator {\color{black}
based on gain--compensated or post--processed data. Post--processing 
is performed using the midpoint correction approach \cite{Jenq}\cite{Frey}\cite{Handel}}.  Results obtained by averaging $3$ records each containing $N=287431$ samples obtained at a nominal 
sine wave frequency of $99.3715$ Hz and sampled at a nominal rate of 
$9135$ samples per second. Input noise standard deviation was estimated as $0.17\Delta$. The ADC transition levels and amplitude gain were estimated in a preceding calibration phase. 
\label{figresults}}
\end{figure}

{\color{black} Processing experimental data highlighted new issues not previously considered during the modeling phase.}
Data were processed both by the LSE method resulting in the 
sine--fit algorithm and by the MVBE.
{\color{black}In some cases also data post--processing was performed before applying the LSE.}
It was observed that:
\begin{itemize}
\item the $3$--parameter sine--fit algorithm  (amplitude, offset and initial record phase) did not perform satisfactorily
because of the uncertainties in the determination of $\lambda$, primarily due to tolerances in the DAQ sampling rate. Thus, a $4$--parameter sine--fit method was used to also estimate the frequency parameter;
\item the performance of the sine--fit algorithm in estimating the sine wave amplitude depended on whether data were or were not corrected for the effect of the ADC gain that was about equal to $1.001$;
{\color{black} since uncompensated data resulted in a worse performance, data were first compensated for the effect of the gain before applying the LSE;}
\item {\color{black} before applying the LSE data could alternatively be post--processed to correct the ADC behavior for the non--uniform distribution of the transition level. This was done by applying the {\em midpoint} correction technique to raw ADC data \cite{Jenq}\cite{Frey}\cite{Handel}. Accordingly, to the $k$--th output code was assigned the value $\frac{1}{2}(T_k+T_{k+1})$, with $T_k, T_{k+1}$ 
being the corresponding code boundary transition levels, so to guarantee that an ideal $45\degree$ line would pass through the centers of all steps in the ADC quantization input--output characteristic \cite{Jenq}.}
\end{itemize}
The performance comparison between estimators is shown in Fig.~\ref{figresults} where the {\em bias}, i.e. the mean error, associated to the usage of both the sine--fit and the MVBE are displayed.
Graphs show that the MVBE almost uniformly reduces the bias even in this case, when the magnitude of the 
INL and DNL of the considered ADC are very small. The difference in performance would be much larger with an ADC with a more severe nonlinear behavior.  
{\color{black}Observe also that the MVBE is even slightly superior over the $4$--parameter sine fit based on {\em midpoint}--corrected data.
The ratio between the squared bias summed over all records and values of $\theta$ is about equal to $0.75$, in favor of the MVBE.
This can be explained by the fact that the midpoint correction is optimal according to the Lloyd's approach \cite{lloyd}, only when the input signal has a PDF 
that is constant within the quantization bin. This happens, for instance, when the input amplitudes are uniformly distributed as when using a deterministic ramp signal \cite{Handel}.
Observe also that, data post--processing using the midpoint correction would not however be able to remove the effects of coarse quantization, resulting in a poor behavior of the LSE, as shown in Fig. \ref{sim2bit}. 
}

\section{Conclusion}
Direct processing of the codes provided by an ADC to estimate parameters associated to ADC input quantities may result in biased estimators, especially when using non uniform quantizers.
In this paper we considered the problem of estimating the amplitude of a sine wave by means of a set of its samples quantized using a non uniform ADCs.
By taking into account knowledge about the actual ADC transition levels it was possible to show that the proposed technique removes most of the bias
associated to the usage of more traditional estimators such as the least{\color{black}s} square one. Both simulation and experimental results were used to 
verify the new estimator properties {\color{black} under several different values of the noise standard deviation}. 
{\color{black} Observe that only overall 
statistical information on the signal amplitudes in the form of a probability density function was considered by the estimator proposed in this paper.
By including also time--related information among input samples 
and thus using knowledge about the correlation of processed data,  further accuracy improvements are expected.}

The idea presented here can be generalized to other types of input signals and suggests that processing ADC output data by also incorporating knowledge about the position of the transition levels provides superior performance over the usage of code--domain only approaches. 

\section*{Acknowledgement}
This work was supported in part by the Fund for Scientific Research (FWO-Vlaanderen), by the Flemish Government (Methusalem), the Belgian Government through the Inter university Poles of Attraction (IAP VII) Program, and by the ERC advanced grant SNLSID, under contract 320378.
% the value of the ADC transition levels. 
\balance

\end{document}